\documentclass[usegraphicx,usenatbib,useAMS]{mn2e}
\usepackage{amsmath}
\usepackage{amssymb}
\usepackage{color}
\usepackage{url}
\usepackage{threeparttable}

\newcommand{\lya}{\ifmmode\mathrm{Ly}\alpha\else{}Ly$\alpha$\fi}
\newcommand{\lyb}{\ifmmode\mathrm{Ly}\beta\else{}Ly$\beta$\fi}
\newcommand{\igm}{\ifmmode\mathrm{IGM}\else{}IGM\fi}
\newcommand{\lae}{\ifmmode\mathrm{LAE}\else{}LAE\fi}
\newcommand{\h}{\ifmmode\mathrm{H}\else{}H\fi}
\newcommand{\hi}{\ifmmode\mathrm{H\,{\scriptscriptstyle I}}\else{}H\,{\scriptsize I}\fi}
\newcommand{\hii}{\ifmmode\mathrm{H\,{\scriptscriptstyle II}}\else{}H\,{\scriptsize II}\fi}
\newcommand{\cmb}{\ifmmode\mathrm{CMB}\else{}CMB\fi}
\newcommand{\qso}{\ifmmode\mathrm{QSO}\else{}QSO\fi}
\newcommand{\eor}{\ifmmode\mathrm{EoR}\else{}EoR\fi}
\newcommand{\cmmc}{\textsc{\small 21CMMC}}
\newcommand{\cmfst}{\textsc{\small 21CMFAST}}
\newcommand{\sense}{{\small 21}\textsc{cmsense}}
\newcommand{\CH}{\textsc{cosmohammer}}
\newcommand{\emcee}{\textsc{emcee}}

\title[21CMMC: astrophysics from the 21 cm \eor{} signal]{21CMMC: an MCMC analysis tool enabling astrophysical parameter studies of the cosmic 21 cm signal}

\author[B. Greig et al.] {Bradley~Greig$^{1}$\thanks{E-mail:~bradley.greig@sns.it} \& Andrei~Mesinger$^{1}$ \\
$^1$Scuola Normale Superiore, Piazza dei Cavalieri 7, I-56126 Pisa, Italy
}

\begin{document}
\maketitle \begin{abstract}
\noindent
We introduce \cmmc{}: a parallelized, Monte Carlo Markov Chain analysis tool,
 incorporating the epoch of reionization (\eor{}) seminumerical simulation \cmfst{}. 
 \cmmc{} estimates astrophysical parameter constraints from 21 cm \eor{} experiments, 
 accommodating a variety of \eor{} models, as well as priors on model parameters and the reionization history.
 To illustrate its utility, we consider two different \eor{} scenarios, one with a single population of galaxies 
 (with a mass-independent ionizing efficiency) and a second, more general model with two different, 
 feedback-regulated populations (each with mass-dependent ionizing efficiencies). 
 As an example, combining three observations ($z=8$, 9 and 10) of the 21 cm power spectrum with a 
 conservative noise estimate and uniform model priors, we find that interferometers with specifications like 
 the Low Frequency Array/Hydrogen Epoch of Reionization Array (HERA)/Square Kilometre Array 1 (SKA1) can constrain common reionization
 parameters: the ionizing efficiency (or similarly the escape fraction), the
 mean free path of ionizing photons and the log of the minimum virial temperature of star-forming haloes to within
 45.3/22.0/16.7, 33.5/18.4/17.8 and 6.3/3.3/2.4 per cent, $\sim1 \sigma$ fractional uncertainty, respectively.
 Instead, if we optimistically assume that we can perfectly characterize the \eor{} modelling uncertainties, we can 
 improve on these constraints by up to a factor of $\sim$few.
 Similarly, the fractional uncertainty on the average neutral fraction can be constrained to within $\lesssim10$ per cent 
 for HERA and SKA1. By studying the resulting impact on astrophysical constraints, \cmmc{} can be used to optimize 
 (i) interferometer designs; (ii) foreground cleaning algorithms; (iii) observing strategies; (iv) alternative statistics 
 characterizing the 21 cm signal; and (v) synergies with other observational programs.
\end{abstract} 

\section{Introduction}
The epoch of reionization (\eor{}) describes the period of enlightenment following the cosmic dark
 ages, when density perturbations grow into the first 
 astrophysical sources (stars and galaxies). These sources produce ultraviolet (UV) 
 ionizing photons, which escape into the intergalactic medium (\igm{}) eventually reionizing the pervasive neutral 
 hydrogen fog.

The \eor{} is rich in astrophysical information, probing the formation and evolution of structure in the Universe,
 the nature of the first stars and galaxies and their impact on the \igm{}
 \citep[see e.g.][]{Barkana:2007p2929,Loeb:2013p2936,Zaroubi:2013p2976}. Currently the 
 astrophysics of the \eor{} are poorly understood. However, a wave of upcoming observations 
 should produce a rapid advance in our understanding. At the forefront of these will be the detection of 
 the redshifted 21 cm spin-flip transition of neutral hydrogen 
 \citep[see e.g.][]{Furlanetto:2006p209,Morales:2010p1274,Pritchard:2012p2958} from several
 dedicated radio interferometers. 

The first-generation 21 cm experiments such as the Low Frequency Array
(LOFAR; \citealt{vanHaarlem:2013p200,Yatawatta:2013p2980})\footnote{http://www.lofar.org/}, the Murchison Wide 
 Field Array (MWA; \citealt{Tingay:2013p2997})\footnote{http://www.mwatelescope.org/} and the 
 Precision Array for Probing the Epoch of Reionization 
 (PAPER; \citealt{Parsons:2010p3000})\footnote{http://eor.berkeley.edu/} are just now coming online.
 These have enabled progress in understanding and characterizing the 
 relevant systematics and foregrounds, and may even yield a 
 detection of the 21 cm power spectrum (PS) during the advanced 
 stages of reionization \citep[e.g.][]{Chapman:2012p344,Mesinger:2014p244}.
  
 In the near future, second-generation experiments, the Square Kilometre Array 
 (SKA; \citealt{Mellema:2013p2975})\footnote{https://www.skatelescope.org} and the Hydrogen Epoch of 
 Reionization Array (HERA; \citealt{Beardsley:2014p1529})\footnote{http://reionization.org} will feature significant increases 
 in collecting area and overall sensitivity. In addition to statistical measurements of the 21 cm PS,
 these second-generation experiments should provide the first tomographic maps of the 21 cm signal from the 
 \eor{} \citep{Mellema:2013p2975}, deepening our understanding of reionization-era physics.
 
However, we are faced with a fundamental question, \textit{what exactly can we learn from these observations?}
Qualitatively, we possess several key insights. Reionization driven by relatively massive galaxies should 
result in larger, more uniform ionized regions \citep[e.g.][]{McQuinn:2007p1665,Iliev:2012p112} 
and the large-scale PS should peak at the 
midpoint of the \eor{} \citep[e.g.][]{Lidz:2008p1744}, unless reionization is driven by X-rays \citep{Mesinger:2013p1835}.
 Abundant, absorption systems should result in an extended reionization 
 \citep[e.g.][]{Ciardi:2006p3030} characterized by small, disjoint ionized regions 
\citep[e.g.][]{Alvarez:2012p1930}. Although these studies provide valuable intuition, they do not 
quantify the constraints and degeneracies among astrophysical parameters.

Therefore, it is imperative to develop an \eor{} analysis tool to quantitatively assess what 
 astrophysical information can be gleaned from the cosmic 21 cm signal. This will serve to guide 
 current and future 21 cm experiments in optimizing their observing strategies and interferometer 
 designs to maximize their \eor{} scientific returns. In recent years, several authors have explored the benefits of this 
 approach by performing \eor{} parameter searches. 
 However, these approaches have been restricted to (i) sampling a finite, fixed grid of either simple analytic models 
 (\citealt{Choudhury:2005p2859,Barkana:2009p116})
 or 3D seminumerical simulations \citep{Mesinger:2012p1131,Zahn:2012p1156, Mesinger:2013p1835}; (ii) performing a 
 Fisher Matrix analysis \citep{Pober:2014p35};  or (iii) Monte Carlo Markov Chain (MCMC) sampling only the 
 reionization history in an analytic framework \citep{Harker:2012p2856,Morandi:2012p2857,Patil:2014p2858} 
 
 Here, we improve on these approaches by developing a public 21 cm \eor{} analysis tool, named \cmmc.
 \cmmc{} uses an optimized version of the well-tested, public simulation tool, \cmfst{} 
 \citep{Mesinger:2007p122,Mesinger:2011p1123}, to generate 3D tomographic maps of the 21 cm brightness 
 temperature, statistically comparing these to a (mock) observation. It quantifies the constraints and degeneracies 
 of astrophysical parameters governing the EoR, within a full Bayesian MCMC framework.

The remainder of this paper is organized as follows. In Section~\ref{sec:21CMMC}, we outline the development
 and implementation of \cmmc. In Section~\ref{sec:SGP}, we highlight its performance at providing astrophysical 
 constraints from a popular theoretical model of the \eor{} containing a single population of ionizing galaxies with a 
 mass-independent ionizing efficiency. To further emphasize the strength and flexibility of 
 \cmmc{} in Section~\ref{sec:DPL}, we then estimate the recovery of astrophysical parameters from a more generalized
 \eor{} model containing two ionizing galaxy populations characterized by their mass-dependent ionizing efficiency. 
 Finally, in Section~\ref{sec:conclusion} we summarize the 
 performance of \cmmc{} and finish with our closing remarks.
 Throughout this work, we adopt the standard set of $\Lambda$CDM cosmological parameters:
 ($\Omega_{\rm m}$, $\Omega_{\rm \Lambda}$, $\Omega_{\rm b}$, $n$, $\sigma_{8}$, $H_{0}$)
 =(0.27, 0.73, 0.046, 0.96, 0.82, 70~${\rm km\, s^{-1} \, Mpc^{-1}}$), measured from the \textit{Wilkinson Microwave Anisotropy Probe} 
 \citep{Bennett:2013p3136} which are consistent with the latest results from 
 Planck \citep{PlanckCollaboration:2014p3099}.

\section{\cmmc{}} \label{sec:21CMMC}

Prior to showcasing the performance of \cmmc{}\footnote{\cmmc{} will be made available at 
http://homepage.sns.it/mesinger/21CMMC.html. Interested users can obtain a pre-release beta version 
by contacting the authors.}, we first outline the various ingredients that contribute to the overall analysis tool.
First, in Section~\ref{sec:21CMFAST}, we summarize the included \eor{} simulation code \cmfst{}. 
We then discuss the likelihood statistic in Section~\ref{sec:likelihood}. Next, in Section~\ref{sec:MCMC} we 
briefly describe the specifics of the MCMC driver and in Section~\ref{sec:TNP} we discuss the telescope
 sensitivities. Finally, in Section~\ref{sec:pipeline} we summarize the full \cmmc{} pipeline.

\subsection{\cmfst{}} \label{sec:21CMFAST}

\cmmc{} employs a streamlined version of \cmfst{}\textunderscore v1.1, a publicly available
semi-numerical simulation code specifically designed for enabling astrophysical parameter searches. 
It employs approximate but efficient methods for \eor{} physics, and is accurate compared to computationally 
 expensive radiative transfer (RT) simulations on scales relevant to 21 cm interferometry, $\geq 1$~Mpc \citep{Zahn:2011p1171}. 
 
Specifically, \cmfst{} generates the \igm{} density, velocity, source and ionization fields by first 
creating a 3D realization of the linear density field within a cubic volume and then evolving the density field using 
the Zel'dovich approximation \citep{Zeldovich:1970p2023} before smoothing on to a lower resolution grid.
 Following this, \cmfst{} estimates the ionization field by comparing the time-integrated number of ionizing photons 
 to the number of baryons within spherical regions of decreasing radius, $R$, following the excursion-set approach 
 described in \citet{Furlanetto:2004p123}. A cell at coordinates $(\boldsymbol{x},z)$ within the simulation 
volume is then tagged as fully ionized if,
\begin{eqnarray}
\label{eq:ioncrit}
\zeta f_{\rm coll}(\boldsymbol{x},z,R,\bar{M}_{\rm min}) \geq 1,
\end{eqnarray}
where $f_{\rm coll}(\boldsymbol{x},z,R,\bar{M}_{\rm min})$ is the fraction of collapsed matter within a spherical 
radius $R$ residing within haloes larger than $\bar{M}_{\rm min}$ 
 \citep{Press:1974p2031,Bond:1991p111,Lacey:1993p115,Sheth:1999p2053}
 and $\zeta$ is an ionizing efficiency describing 
the conversion of mass into ionizing photons (see Section~\ref{sec:Zeta}). Partial ionization is additionally included 
for cells not fully ionized by setting the ionized fraction of a cell smoothed at the minimum smoothing scale, 
$R_{\rm cell}$ to $\zeta f_{\rm coll}(\boldsymbol{x},z,R_{\rm cell},\bar{M}_{\rm min})$.

It is common to characterize the \eor{} with just three parameters: (i) the ionizing efficiency of high-$z$ galaxies; 
(ii) the mean free path of ionizing photons; (iii) the minimum virial temperature hosting star-forming galaxies.
These parameters are somewhat empirical and in most cases are averaged over redshift and/or halo mass dependences.
Nevertheless, they offer the flexibility to describe a wide variety of EoR signals, and can have a straightforward physical interpretation.
Below, we summarize each of these three parameters and their physical origins.

\subsubsection{Ionizing efficiency, $\zeta$} \label{sec:Zeta}

The ionizing efficiency of high-$z$ galaxies (equation~\ref{eq:ioncrit}) can be expressed as 
 \begin{eqnarray} \label{eq:Zeta}
 \zeta = 30\left(\frac{f_{\rm esc}}{0.3}\right)\left(\frac{f_{\star}}{0.05}\right)\left(\frac{N_{\gamma}}{4000}\right)
 \left(\frac{2}{1+n_{\rm rec}}\right)
 \end{eqnarray}
 where, $f_{\rm esc}$ is the fraction of ionizing photons escaping into the IGM, $f_{\star}$ is the fraction of 
 galactic gas in stars, $N_{\gamma}$ is the number of ionizing photons produced per baryon in stars and 
 $n_{\rm rec}$ is the typical number of times a hydrogen atom recombines. While our reionization model only depends 
 on the product of equation~(\ref{eq:Zeta}), we provide reasonable estimates for the individual factors. The choice of 
 $N_{\gamma}\approx4000$ is expected from Population II stars \citep[e.g.][]{Barkana:2005p1934}, while $f_{\star}$ and 
 $f_{\rm esc}$ are extremely uncertain in high-$z$ galaxies 
 \citep[e.g.][]{Gnedin:2008p1935,Wise:2009p1968,Ferrara:2013p1248}, though our choices are consistent with 
 observations of galaxy luminosity functions \citep[e.g.][]{Robertson:2013p2123,Cooke:2014p2861,Dayal:2014p2876}.
 Finally, $n_{\rm rec} \sim 1$ is consistent with the models of \citep{Sobacchi:2014p1157} which result in a `photon-starved' end to reionization, consistent with emissivity estimates from the \lya{} forest \citep[e.g.][]{Bolton:2007p3273,McQuinn:2011p3293}.
 We allow $\zeta$ to vary within the range $\zeta\in[5,100]$. Since, $\zeta$ is empirically defined unlike $f_{\rm esc}$ which can be 
 observationally constrained at lower redshift, we will often provide both when recovering our parameter constraints. 
 For our range of $\zeta$, this corresponds to $f_{\rm esc}\in[0.05,1]$ using our fiducial values adopted in 
 equation~(\ref{eq:Zeta}). In this work, we will first consider a constant $\zeta$ (Section~\ref{sec:SGP}) before considering 
 a more generalized halo mass-dependent model in Section~\ref{sec:DPL}.

\subsubsection{Mean free path of ionizing photons within ionized regions, $R_{\rm mfp}$} 

The propagation of ionizing photons through the \igm{} strongly depends on the abundances and properties 
of absorption systems (Lyman limit as well as more diffuse systems), which are below the resolution limits of 
EoR simulations. These systems act as photon sinks, roughly dictating the maximum scales to which \hii{}
 bubbles can grow around ionizing galaxies.  In EoR modelling this effect is usually parametrized with an implicit 
 maximum horizon for ionizing photons (corresponding to the maximum filtering scale in excursion-set EoR models). 
 Physically, this maximum horizon is taken to correspond to the mean free path of ionizing photons within ionized regions, 
 $R_{\rm mfp}$ \citep[e.g.][]{OMeara:2007p3360,Prochaska:2009p3339,Songaila:2010p3348,McQuinn:2011p3293}.
 Motivated by recent sub-grid recombination models \citep{Sobacchi:2014p1157}, here we explore the range 
 $R_{\rm mfp}\in[5, 20]$~cMpc\footnote{The ionization structure of these subgrid models of recombinations
 inside unresolved systems does not in fact directly translate to a constant, uniform value of $R_{\rm mfp}$.  These authors find 
 that it takes a long time for cosmic \hii{} regions to approach the Str{\"o}mgren limit, with their expansion only slowing more and more as reionization progresses.  In other words, the mean free path (which is only defined in terms of the {\it instantaneous} recombination rate) is almost always {\it larger} than the actual size of the \hii{} regions (which depend on the {\it cumulative} number of recombinations), even at the late stages of reionization.
 The resulting ionization fields have a dearth of ionization structure on scales smaller than the instantaneous mean free path. 
 Moreover, the spatial correlations between the sources and sinks of ionizing photons results in sizeable spatial and
 temporal variations in the mean free path. Nevertheless, an effective, uniform maximum horizon for ionizing photons of $\sim 10$ Mpc can reproduce the 21 cm 
 power spectra of the \citet{Sobacchi:2014p1157} simulations to within tens of percent on relevant scales, at the epochs studied in that work 
 $0.25 \lesssim \bar{x}_{\hi{}} \lesssim 0.75$ (Sobacchi, private communication; Mesinger \& Greig, in preparation).  For historical context, we use the variable `$R_{\rm mfp}$' to denote this effective horizon set by subgrid recombinations, but caution that the actual mean free path is larger than this parameter, as discussed above. }. While this range is relatively narrow, we point out that this parameter only becomes 
 important once the size of the ionized regions become larger than $R_{\rm mfp}$ 
 \citep[e.g.][]{McQuinn:2007p1665,Alvarez:2012p1930,Mesinger:2012p1131}. Expanding the allowed range of 
 $R_{\rm mfp}$ we consider in this work would therefore not greatly modify our conclusions.

\subsubsection{Minimum virial temperature of star-forming haloes, $T^{\rm Feed}_{\rm vir}$} \label{sec:Mthresh}

Throughout, we choose to define the minimum threshold for a halo hosting a star-forming galaxy to be in terms of 
its virial temperature, which regulates processes important for star formation: gas accretion, cooling and retainment 
of supernovae outflows. The virial temperature is related to the halo mass via, \citep[e.g.][]{Barkana:2001p1634}
\begin{eqnarray}
M_{\rm min} &=& 10^{8} h^{-1} \left(\frac{\mu}{0.6}\right)^{-3/2}\left(\frac{\Omega_{\rm m}}{\Omega^{z}_{\rm m}}
\frac{\Delta_{\rm c}}{18\upi^{2}}\right)^{-1/2} \nonumber \\
& & \times \left(\frac{T_{\rm vir}}{1.98\times10^{4}~{\rm K}}\right)^{3/2}\left(\frac{1+z}{10}\right)^{-3/2}M_{\sun},
\end{eqnarray}
where $\mu$ is the mean molecular weight, $\Omega^{z}_{\rm m} = \Omega_{\rm m}(1+z)^{3}/[\Omega_{\rm m}(1+z)^{3} + 
\Omega_{\Lambda}]$, and $\Delta_{\rm c} = 18\upi^{2} + 82d - 39d^{2}$ where $d = \Omega^{z}_{\rm m}-1$. 
Typically, $T_{\rm vir}\approx10^{4}$~K has been adopted in the literature as the minimum temperature when efficient 
atomic cooling occurs, corresponding to a DM halo mass of $\approx10^{8}M_{\sun}$ at $z\sim10$. In principle, this could
be lower as the first stars are likely hosted within haloes with $M_{\rm halo}\approx 10^{6}$--$10^{7}$~$M_{\sun}$ 
\citep{Haiman:1996p2144,Abel:2002p2149,Bromm:2002p2153}. However,
star formation within these haloes is likely inefficient (a few stars per halo) and can be quickly ($z>20$)
suppressed by Lyman--Werner or other feedback processes well before the \eor{}
\citep{Haiman:2000p2155,Ricotti:2001p2160,Haiman:2006p2169,Mesinger:2006p2171,Holzbauer:2012p2890,Fialkov:2013p2903}.

Here, we allow for a critical temperature threshold, $T^{\rm Feed}_{\rm vir}$, which can even be larger than $10^{4}$~K. Below 
$T^{\rm Feed}_{\rm vir}$ (but above $10^{4}$~K), haloes are still sufficiently large to produce low-mass galaxies; 
however, their potential wells are not deep enough to retain supernovae-driven outflows that quench star formation and 
their corresponding contribution to reionization \citep[e.g.][]{Springel:2003p2176}. In this work, we allow 
$T^{\rm Feed}_{\rm vir}$ to vary within the range $T^{\rm Feed}_{\rm vir}\in[10^{4},2\times10^{5}]$~K, with the lower limit
set by the atomic cooling threshold and the upper limit consistent with observed high-$z$ Lyman break galaxies 
(see for example Fig.~\ref{fig:UVLum}).

\subsection{Likelihood statistic} \label{sec:likelihood}

The offset of the 21 cm brightness temperature relative to the 
\cmb{} temperature, $T_{\gamma}$ \citep[e.g.][]{Furlanetto:2006p209}, can be written as,
\begin{eqnarray}
\delta T_{\rm b}(\nu) &=& \frac{T_{\rm S} - T_{\gamma}}{1+z}\left(1-{\rm e}^{-\tau_{\nu_0}}\right) \nonumber \\
&\approx& 27x_{\hi{}}(1+\delta_{\rm nl})\left(\frac{H}{{\rm d}v_{\rm r}/{\rm d}r+H}\right)
\left(1 - \frac{T_{\gamma}}{T_{\rm S}}\right) \nonumber \\
& & \times \left(\frac{1+z}{10}\frac{0.15}{\Omega_{\rm m}h^{2}}\right)^{1/2}
\left(\frac{\Omega_{\rm b}h^{2}}{0.023}\right)~{\rm mK},
\end{eqnarray}
where $T_{\rm S}$ is the gas spin temperature, $\tau_{\nu_0}$ is the optical depth at the 21 cm frequency, 
$\nu_0$, $\delta_{\rm nl}(\boldsymbol{x},z)$ is the evolved (Eularian) overdensity, $H(z)$ is the Hubble parameter, 
${\rm d}v_{\rm r}/{\rm d}r$ is the gradient of the line of sight component of the velocity and all quantities are evaluated at 
redshift $z = \nu_{0}/\nu - 1$. Throughout this work, we include the effects of peculiar velocities and 
assume $T_{\rm S} \gg T_{\gamma}$ (likely appropriate for the
advanced stages of reionization, corresponding to an average neutral fraction of $\bar{x}_{\hi} \lesssim 0.8$, 
e.g.\ \citealt{Mesinger:2013p1835}) as the explicit computation of the spin temperature within \cmfst{} is 
computationally expensive; however, in future we will remove this assumption. 

Although \cmfst{} produces 3D maps of the 21 cm brightness temperature, we require a goodness of fit statistic in order
to quantitatively compare an \eor{} realization with the mock observation. We choose the common spherically averaged 21 cm PS, 
$\Delta^{2}_{21}(k,z) \equiv k^{3}/(2\upi^{2}V)\,\delta \bar{T}^{2}_{\rm b}(z)\,\langle |\delta_{21}(\boldsymbol{k},z)|^{2}\rangle_{k}$
 where $\delta_{21}(\boldsymbol{x},z) \equiv \delta T_{\rm b}(\boldsymbol{x},z)/\delta \bar{T}_{\rm b}(z) -1$, 
 to be this statistic. It is important to note that while we restrict our default analysis to 
 $\Delta^{2}_{21}(k,z)$, in principle any statistical measure of the cosmic 21 cm signal could be used in our framework.

\subsection{The MCMC driver} \label{sec:MCMC}

An MCMC based \eor{} tool is a computationally challenging prospect since each step in the computation chain
 requires a \cmfst{} realization. For reference, computing a 128$^{3}$ 21 cm realization at a 
 single redshift takes $\sim3$s using a single core. This is efficient, however, it is a case of diminishing 
 returns if we attempt to improve the computation time using the inbuilt multithreading of \cmfst{}. 
 Therefore, constructing an MCMC analysis 
 tool which samples \cmfst{} utilizing existing serial MCMC algorithms such as Metropolis--Hastings 
 (\citealt{Metropolis:1953p2224,Hastings:1970p97}), which require multiple long 
 computation chains, would be too computationally inefficient.

In recent years, several new, more efficient and parallel MCMC algorithms have been developed. Once such 
 algorithm is the affine invariant ensemble sampler developed by \citet{Goodman:2010p843}. This algorithm has since
 been modified and improved before being released as the publicly available \textsc{python} module \emcee{} by 
 \citet{ForemanMackey:2013p823}. Recently, 
 \citet{Akeret:2012p842} discussed the virtues of adopting these computational algorithms for applications of
 cosmological parameter sampling. These authors released an easy to implement, massively parallel, publicly available 
 \textsc{python} module \CH{} which has already been extensibly adopted for a wide variety of cosmological application.  
 We therefore choose to build \cmmc{} using a modified version of \CH{} which performs 
 our MCMC reionization parameter sampling by calling \cmfst{} at each step 
 in the computation chain to compute the log-likelihood.

\subsection{Telescope noise profiles} \label{sec:TNP}

To illustrate the performance of \cmmc{} for current and future 21 cm experiments, we compute the telescope 
sensitivities using the \textsc{python} module \sense{}\footnote{https://github.com/jpober/21cmSense}\citep{Pober:2013p41,Pober:2014p35}. 
 In this section, we outline only the specifics and assumptions we make in producing our realistic telescope noise profiles,
 summarizing the key telescope parameters in Table~\ref{tab:TelescopeParams}, and defer the reader to the
 more detailed discussions within \citet{Parsons:2012p95} and \citet{Pober:2013p41,Pober:2014p35}.

The thermal noise PS is computed at each $uv$-cell  according to the following 
\citep[e.g.][]{Morales:2005p1474,McQuinn:2006p109,Pober:2014p35},
\begin{eqnarray} \label{eq:NoisePS}
\Delta^{2}_{\rm N}(k) \approx X^{2}Y\frac{k^{3}}{2\upi^{2}}\frac{\Omega^{\prime}}{2t}T^{2}_{\rm sys},
\end{eqnarray} 
where $X^{2}Y$ is a cosmological conversion factor between observing bandwidth, frequency and comoving distance
 units, $\Omega^{\prime}$ is a beam-dependent factor derived in \citet{Parsons:2014p781}, $t$ is the total time spent by 
 all baselines within a particular $k$ mode and $T_{\rm sys}$ is the system temperature, the sum of the receiver 
 temperature, $T_{\rm rec}$, and the sky temperature $T_{\rm sky}$. For all telescope configurations considered in this work, 
 we assume drift scanning with a total synthesis time of 6~h per night.  While HERA can only operate in drift scan mode, 
 LOFAR and SKA can perform tracked scans.  In the future, we will investigate various observing strategies; however, drift 
 scanning suffices here for our like-to-like comparisons.
 
Of great concern for 21 cm experiments is the estimation and removal of the bright foreground emission. Most notable of 
 these are the chromatic effects due to how the interferometric array's $uv$ coverage depends on frequency. 
 It has been shown that these effects do not lead to mode-mixing across all frequencies, but instead 
 localize the majority of the foreground noise into a foreground `wedge' within cylindrical 2D $k$-space 
 \citep{Datta:2010p2792,Morales:2012p2828,Parsons:2012p2833,Trott:2012p2834,Vedantham:2012p2801,Thyagarajan:2013p2851,Liu:2014p3465,Liu:2014p3466}. 
 This results in a relatively pristine observing window where the cosmological 21 cm signal is dominated only by thermal noise.
 However, it is still uncertain where to define the transition from the foreground `wedge' to the \eor{} observing window. 

 Moreover, the summation over redundant baselines within an array configuration can reduce
 uncertainties in the thermal noise estimates \citep{Parsons:2012p95}. However, the gain
 depends on whether they are perfectly or partially coherent \citep{Hazelton:2013p1481}.
 
 In \citet{Pober:2014p35}, these authors consider numerous foreground removal strategies, including a variety of 
 treatments for the summation over redundant baselines and how to define the transition from the foreground `wedge'. 
 We defer the reader to their work for more detailed discussions, and highlight that throughout this work, we adopt what they refer
 to as their `moderate' foreground removal. For this, we assume a fixed location of the 
 foreground `wedge' to extend $\Delta k_{\parallel} = 0.1 \,h$~Mpc$^{-1}$ beyond the horizon limit \citep{Pober:2014p35} and
 compute our estimates of the thermal noise PS sampling only $k$-modes that fall within the \eor{} observing window.
 This strategy further assumes the coherent addition of all redundant baselines.
 
 \begin{table}
\begin{tabular}{@{}lcccc}
\hline
Parameter & LOFAR & HERA & SKA \\
\hline
Telescope antennas & 48 & 331 & 866 \\
Diameter (m) & 30.8 & 14 & 35 \\
Collecting area (m$^2$) & 35\,762 & 50\,953 & 833\,190\\
$T_{\rm rec}$ (K) & 140 & 100 & 0.1T$_{\rm sky}$ + 40 \\
Bandwidth (MHz) & 8 & 8 & 8 \\
Integration time (hrs) & 1000 & 1000 & 1000 \\
\hline
\end{tabular}
\caption{Summary of telescope parameters we use to compute sensitivity profiles (see the text for further details).}
\label{tab:TelescopeParams}
\end{table}
 
 We include another level of conservatism to our errors, applying a strict $k$-mode cut to our 21 cm PS.  We sample only 
 modes above $|k|$ = 0.15~Mpc$^{-1}$. Following the moderate foreground removal strategy of \citet{Pober:2014p35}, it 
 effectively turns out that there is limited information available below this $k$-mode; however, we still apply this strict cut on 
 top of the estimated total noise PS. Modes with $|k| \geq$ 0.15~Mpc$^{-1}$ can still be affected by the 
 foreground `wedge' due to its spherically asymmetric structure, hence we apply both cuts.

Since equation~(\ref{eq:NoisePS}) is computed for each individual $k$-mode, the sample variance of the 
 cosmological PS can be easily included to produce an estimate of the total noise
 by performing an inverse-weighted summation over all the individual modes \citep{Pober:2013p41},
\begin{eqnarray} \label{eq:T+S}
\delta\Delta^{2}_{\rm T+S}(k) = \left(\sum_{i}\frac{1}{(\Delta^{2}_{{\rm N},i}(k) + \Delta^{2}_{21}(k))^{2}}\right)^{-1/2},
\end{eqnarray}
where $\delta\Delta^{2}_{\rm T+S}(k)$ is the total uncertainty from thermal noise and sample variance in a given 
 $k$-mode and $\Delta^{2}_{21}(k)$ is the cosmological 21 cm PS. Here, we assume Gaussian errors for the cosmic-variance term, which is a good approximation on large scales.
 
Within this work, we restrict our analysis to three telescope designs (LOFAR, HERA and the SKA).
 In \citet{Pober:2014p35}, these authors qualitatively compare the merits of a variety of telescope designs through a 
  measurement of the overall sensitivity of the 21 cm PS. To perform this direct comparison, these authors 
  set a variety of design specific parameters to be equivalent. Instead, we perform our comparison using the exact design 
  specifics of each experiment as outlined below. Note, the difference is
  marginal, however using system specific parameters we better mimic the expected overall sensitivities of 
  the modelled telescope. For all, we model $T_{\rm sky}$ using the frequency dependent 
  scaling $T_{\rm sky} = 60\left(\frac{\nu}{300~{\rm MHz}}\right)^{-2.55}~{\rm K}$ \citep{Thompson2007}.
 
 \begin{itemize}
\item[1.] \textit{LOFAR:} we model LOFAR using the antennas positions listed in \citet{vanHaarlem:2013p200}.
 Following the approach of \citet{Pober:2014p35}, we focus solely on the Netherlands core for the purposes of detecting
 the 21 cm PS assuming that all 48 high-band antennas (HBA) stations can be correlated separately 
 maximizing the total number of redundant baselines. We assume $T_{\rm rec} =140$~K 
 consistent with \citet{Jensen:2013p1389}.
\\
\item[2.] \textit{HERA:} we follow the design specifics outlined in \citet{Beardsley:2014p1529}, 
 where they propose a final design including 331 antennas. Each antennas is 14m in diameter
 closely packed into a hexagonal configuration to maximize the total number of redundant 
 baselines \citep{Parsons:2012p95}. While the total observing area for HERA is an order of magnitude lower than 
 the SKA, its design is optimized specifically for PS measurements allowing for comparable sensitivity
\citep{Pober:2014p35}\footnote{These
authors found a factor of 2.5 difference between their SKA and HERA designs for their estimated total sensitivity 
on a measurement of the 21 cm PS at $z=9.5$. However, this factor was a result of a numerical error in their 
calculation of their SKA sensitivities. Accounting for this, their HERA design has in fact only a marginally 
reduced sensitivity relative to the SKA for their pessimistic and moderate foreground scenarios
 (Pober, private communication)}. However, our 331 antennas telescope is smaller than the larger 
 547 antennas instrument proposed by these authors, therefore, our design performs more as an 
intermediate instrument. For HERA, we model the total system temperature as $T_{\rm sys} = 100 + T_{\rm sky}~{\rm K}$.
 \\
\item[3.] \textit{SKA:} we mimic the current design brief for the SKA-low Phase 1 instrument outlined in the 
 SKA System Baseline Design document\footnote{http://www.skatelescope.org/wp-content/uploads/2012/07/SKA-TEL-SKO-DD-001-1\textunderscore BaselineDesign1.pdf}. Specifically, SKA-low Phase 1 includes a total of 911 
 35 m antennas stations, with 866 stations normally distributed in radius within a core extending out to a radius of 
 $\sim$2 km and a further 45 stations arranged into three spiral arms each with 15 stations extending to 
 $\sim50$~km. For our SKA design we include only the core stations, as the spiral arms add little to the PS 
 sensitivity. As in \citet{Pober:2014p35}, we model the central core of SKA as a hexagon to maximize baseline redundancy, 
 however our core extends out to a radius of $\sim300$m which includes a total of 218 telescope antennas. We then
 randomly place the telescope antennas normalising to ensure we contain 433 and 650 telescopes within a radii of 
 600 and 1000 m and randomly place the remaining 216 antennas at a radii larger than 1000 m. 
 The total SKA system temperature is modelled as outlined in the SKA System Baseline Design,
 $T_{\rm sys} = 1.1T_{\rm sky} + 40~{\rm K}$.

\end{itemize}

\subsection{The \cmfst{} pipeline} \label{sec:pipeline}

Eventually \cmmc{} will be performed on actual observations.  However for this proof-of-concept, we first create mock 
observations of the cosmic \eor{} signal and generate sensitivities for various instruments\footnote{At present, \cmmc{}
computes only co-evolving data cubes of the \eor{}. In future, we intend to improve the simulation pipeline to mimic a more
realistic observational pipeline, taking into account several effects such as finite light-cones, foreground cleaning etc.}. 
Using \cmfst{}, we perform a 
high-resolution, large-box realization of our fiducial \eor{} model, simulated within a 500$^{3}$~cMpc$^{3}$ box with an 
initial grid of 1536$^{3}$ voxels smoothed on to a 256$^{3}$ grid. We then calculate the 21 cm PS from this simulation at 
several different redshifts and compute the total noise contribution from thermal and sample variance using \sense{}.
In addition to the thermal noise and sample variance, we add an additional uncertainty due to EoR modelling, accounting for, e.g. 
differences in the implementation of RT \citep[e.g.][]{Iliev:2006p3393} and errors in semi-numerical 
approximations \citep[e.g.][]{Zahn:2011p1171}.  In other words, even if we have a well behaved Universe characterized exactly 
by our 3 or 5 parameter models, the simulated PS will be different from the true one due to the numerical 
implementation. We assume for this a constant multiplicative error of 25 per cent, guided by the observed 
10--30 per cent difference at the relevant scales between \cmfst{} and the cosmological RT simulations in \citet{Zahn:2011p1171}. For each mock observation of the 21 cm PS, we add this 25 per cent error in quadrature with the noise contribution from equation~(\ref{eq:T+S}). In principle this would be a correlated error, 
however, at this stage a constant multiplicative factor is sufficient.  Below, we include some results excluding this error, showing the 
maximum information which can be extracted from the signal assuming a perfect characterization of modelling uncertainties.
 
To sample the reionization parameter space in \cmmc, we perform a smaller, lower resolution \eor{} simulation. 
Using a different set of initial conditions to the mock observations, we simulate a 250$^{3}$~cMpc$^{3}$ volume with 
 an initial grid of 768$^{3}$ voxels smoothed down on to a 128$^{3}$ grid. Our two simulations are both sampled at the 
 same physical resolution ($\sim2$~Mpc), and resampled by the same factor of 6 from the initial density grid to 
 ensure convergence between the recovered 21 cm PS from the mock observation and the PS from the smaller 
 128$^{3}$ simulations. To account for the impact of shot noise on the smallest spatial modes (largest $k$) within 
 these low-resolution simulations, we apply an additional cut at $k\geq1.0~{\rm Mpc}^{-1}$. 
 
At each step in the MCMC algorithm we compute the 21 cm PS corresponding to that point in our astrophysical
 parameter space using the 3D realization from \cmfst{}. We compute its likelihood using the $\chi^{2}$ 
 statistic between $k=0.15$ and 1.0~Mpc$^{-1}$ to assess whether to accept or reject the new proposed set of \eor{} parameters.
 
Finally, we briefly discuss the improvements in our method to sample the astrophysical \eor{} parameter space, relative to 
previous works. These previous approaches have been restricted to either sampling a fixed, course grid of 
reionization models \citep[e.g.][]{Barkana:2009p116, Mesinger:2012p1131,Zahn:2012p1156} 
or the fundamental assumption of Gaussian errors in Fisher Matrix applications (\citealt{Pober:2014p35}; although 
new methods exist to overcome this assumption, e.g.\ \citealt{Joachimi:2011p3193,Sellentin:2014p3227}). 
Instead, with a Bayesian MCMC framework, we are able to more efficiently sample the \eor{} parameter space and by 
construction obtain direct estimates of the individual probability distribution functions (PDFs) of the \eor{} parameters, 
removing the \textit{a priori} assumptions on the intrinsic probability distribution (i.e.\ Gaussian). 
Moreover, by sampling over 10$^{5}$ realizations within \cmmc{} this allows us to sample many features
within the \eor{} parameter space not visible in course-grid models. Additionally, this allows for the 
better characterization of parameter degeneracies, which is a problem for Fisher Matrix analyses.

\section{Single Ionizing Galaxy Population} \label{sec:SGP}

To illustrate the use of \cmmc, we first adopt the popular three-parameter EoR model, discussed above.
This model is characterized by a single population of efficient star-forming galaxies hosted by haloes above our defined
critical temperature $T^{\rm Feed}_{\rm vir}$. The ionizing efficiency of this model is a step function,
\begin{eqnarray} \label{eq:zeta_0}
\zeta(T_{\rm vir}) = \left\{ \begin{array}{cl} 
      \zeta_{0}  & {T_{\rm vir} \geq T^{\rm Feed}_{\rm vir}}\\
      0 & {T_{\rm vir} < T^{\rm Feed}_{\rm vir}}.
\end{array} \right.
\end{eqnarray}
In this model, a constant fraction of the host halo mass goes into the production of ionizing photons. 
To illustrate the performance of \cmmc, we choose $\zeta_{0} = 30$, $R_{\rm mfp} = 15$~Mpc 
and $T^{\rm Feed}_{\rm vir} = 3\times10^{4}$~K for our mock observation. Our 
chosen values are roughly consistent with observational data, though they merely serve for illustrative purposes.

\subsection{Building physical intuition} \label{sec:intuition}

\begin{figure} 
	\begin{center}
		\includegraphics[trim = 0.5cm 1cm 0cm 1cm, scale = 0.46]{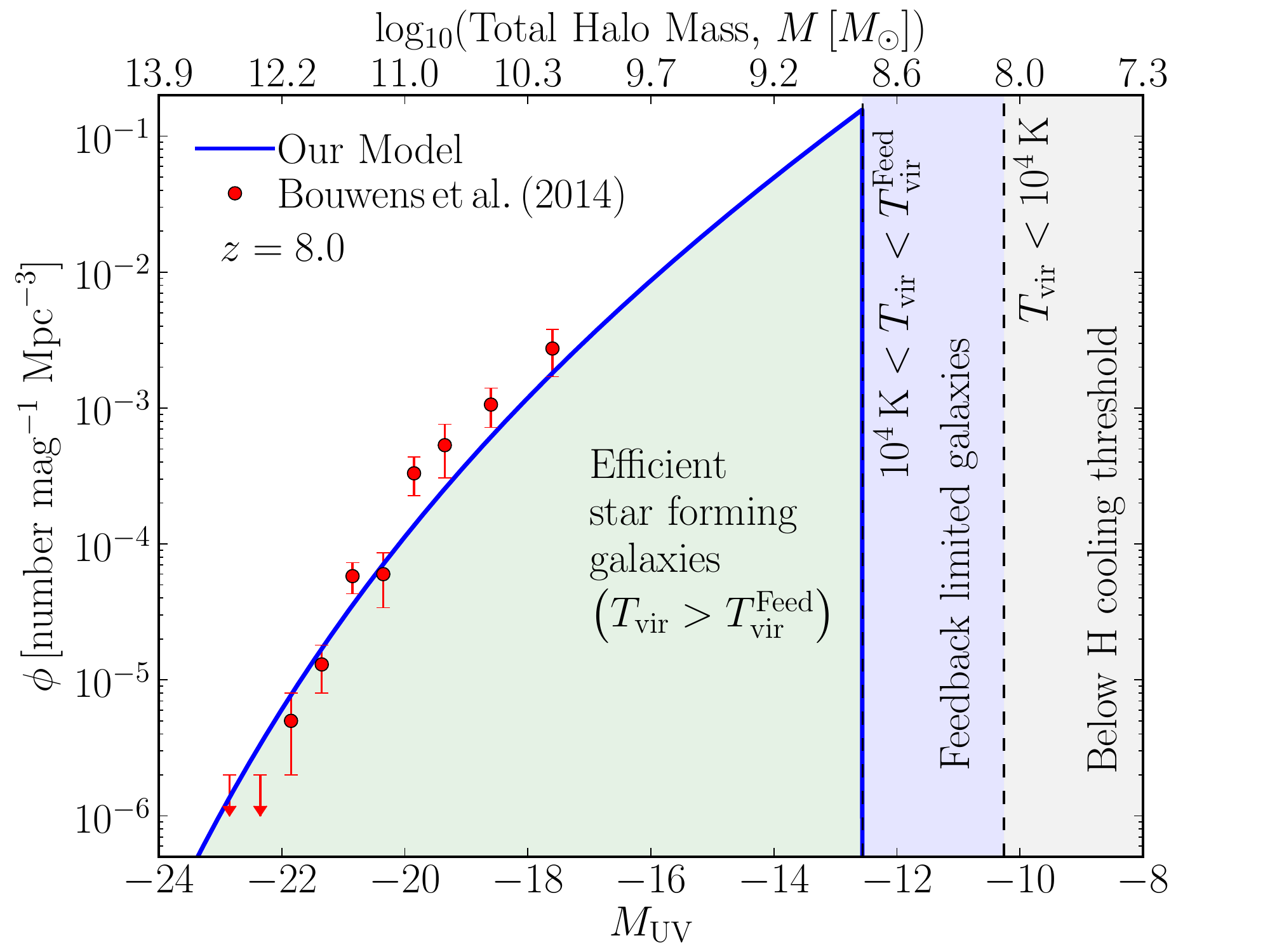}
	\end{center}
\caption[]
{A schematic picture of the $z=8$ non-ionizing UV luminosity function (blue curve) from our fiducial 
single population reionization model (used for the mock observation) characterized 
by a threshold temperature of $T^{\rm Feed}_{\rm vir}=3\times10^{4}\,{\rm K}$ ($M\sim10^{8.78}\,M_{\sun}$~at $z=8$). 
Haloes above $T^{\rm Feed}_{\rm vir}$ host
typical star-forming galaxies contributing to the reionization of the IGM (green shaded region). Below 
$T^{\rm Feed}_{\rm vir}$, haloes are small enough to be affected by internal feedback 
processes (blue shaded region). At $T_{\rm vir}<10^{4}\,{\rm K}$~
($M\sim10^{8.06}\,M_{\sun}$~at $z=8$), gas can no longer efficiently cool to accrete on to haloes. Note that the model curve
at $T_{\rm vir} > T^{\rm Feed}_{\rm vir}$ is obtained from a power-law fit to the halo mass as a function of UV magnitude recovered 
from abundance matching, which we can use to imply that for our fiducial choice of a constant
ionizing efficiency we roughly obtain the scaling, $f_{\rm esc}\propto M^{0.32}$
(see the text for further discussions).}
\label{fig:UVLum}
\end{figure}

\subsubsection{UV luminosity function} \label{sec:UVLum}

It is first constructive to illustrate how this \eor{} model would appear with respect to current observations. 
In Fig.~\ref{fig:UVLum}, we provide a schematic picture of our single
 ionizing galaxy population compared to the observed non-ionizing ($\sim$ 1500 \AA) UV luminosity function at $z=8$
 of \citet{Bouwens:2014p1528}.  To obtain the UV luminosity of our galaxy population, we use the 
 Schechter function parameter values described in \citet{Kuhlen:2012p1506} and abundance matching to estimate the 
 host halo mass as a function of the UV magnitude, $M(M_{\rm UV})$. Fitting this expression as a power-law, we convert
 our $T^{\rm Feed}_{\rm vir}$ into a UV magnitude which produces a sharp drop 
 in the UV luminosity function at $M_{\rm UV} = -12.6$ ($M\sim10^{8.78}\,M_{\sun}$). By construction, the non-ionizing UV luminosity 
 function matches the observational data and we extrapolate down into the faint-end until $T^{\rm Feed}_{\rm vir}$, below which 
 no ionizing photons are produced/escape the galaxy. 
 In our illustrative model, this drop, tracing fundamental galaxy formation physics, is well beyond the sensitivity
 limits even for JWST \citep[e.g.][]{Salvaterra:2011p2908}, showcasing the power of our approach.
 
The UV luminosity function with an appropriate dust model can constrain the star formation rate, which 
(approximately) collapses the uncertainty in the ionizing emissivity to just that in the escape fraction, $f_{\rm esc}$.
In other words, the total ionizing luminosity,
$L_{\rm ion}$ is roughly proportional to the product of the non-ionzing UV luminosity ($L_{\rm UV}$) and $f_{\rm esc}$:
$L_{\rm ion}(M) \propto f_{\rm esc}(M)L_{\rm UV}(M)$.
Our fiducial choice of a constant $\zeta$ implies $L_{\rm ion}\propto M$, from which we can obtain an expression for 
$L_{\rm UV}(M)$ using our power-law expression for $M(M_{\rm UV})$ to recover $L_{\rm UV}(M)\propto M^{0.68}$ for the faint end
of the luminosity function. 
Therefore, our illustrative, single population model 
at $z=8$ roughly translates to $f_{\rm esc}\propto M^{0.32}$.
 
  \begin{figure*} 
	\begin{center}
		\includegraphics[trim = 0.3cm 0.5cm 0cm 0.2cm, scale = 0.9]{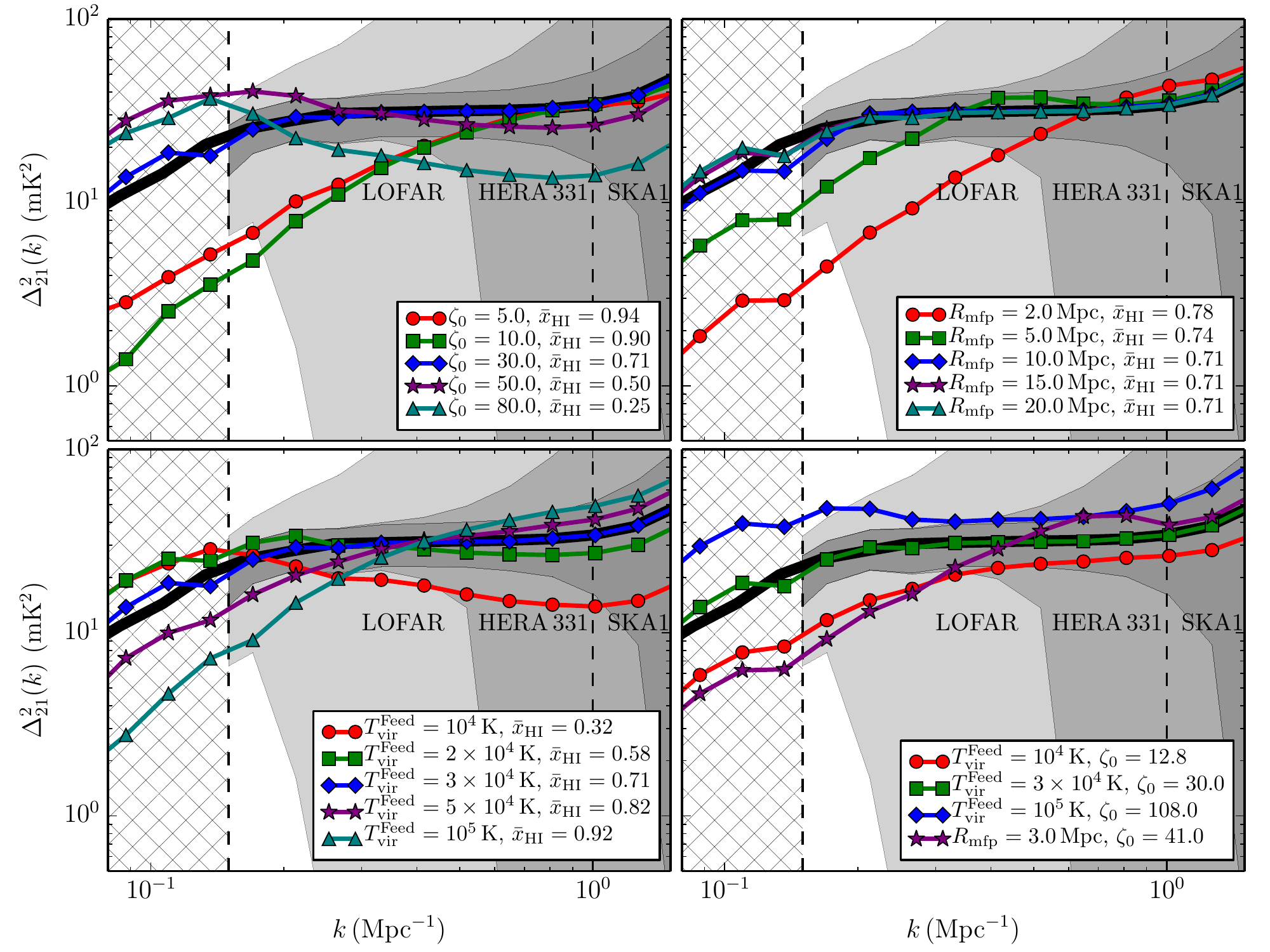}
	\end{center}
\caption[]
{The 21 cm PS at $z=9$ for various values of our three 
reionization parameters, $\zeta_{0}$, $R_{\rm mfp}$ and $T^{\rm Feed}_{\rm vir}$. In all panels, the thick 
black curve corresponds to our fiducial 
reionization model, with ($\zeta_{0}$, $R_{\rm mfp}$, $T^{\rm Feed}_{\rm vir}$) = (30, 15~Mpc, $3\times10^{4}$~K) 
and a neutral fraction of $\bar{x}_{\hi{}} = 0.71$. Light to dark shaded regions correspond to the 1$\sigma$ errors 
of the total noise PS (see Sections~\ref{sec:TNP} and~\ref{sec:pipeline}) for the three 21 cm experiments 
we consider in this analysis (LOFAR, HERA and SKA), including a 25 per cent modelling uncertainty. 
Dashed vertical lines at $k=0.15$ and $k=1.0$~Mpc$^{-1}$ demarcate the physical scales on which we choose to fit the 
21 cm PS, while the hatched region corresponds to the foreground dominated region. We show the impact of the ionizing 
efficiency $\zeta_{0}$ (top left), the maximum horizon for ionizing photons, $R_{\rm mfp}$ (top right) and the minimum virial 
temperature of star-forming galaxies, $T^{\rm Feed}_{\rm vir}$ (bottom left). Bottom right: the 21 cm PS corresponding to 
different EoR parameters, but at the same EoR epoch ($\bar{x}_{\hi{}}$) as the fiducial simulation.}
\label{fig:3param_variation}
\end{figure*}  
   
\subsubsection{How does the 21 cm PS depend on the \eor{} parameters?}   \label{sec:variation}

In Fig.~\ref{fig:3param_variation}, we highlight how each individual \eor{} parameter affects the 21 cm PS, 
and also provide a visual representation of the total noise profiles for each telescope. In all panels, we show the 
mock $z=9$ observation for our fiducial \eor{} parameters (solid black curve) which results in $\bar{x}_{\hi{}} = 0.71$.
In order to facilitate the discussion, we decompose the 21 cm PS to first order (neglecting the contribution from redshift-space 
distortions and spin temperature fluctuations),
\begin{eqnarray} \label{eq:decomp}
\Delta^{2}_{21}(k) &\propto& \bar{x}^{2}_{\hi{}}\Delta^{2}_{\delta\delta}(k) -2\bar{x}_{\hi}\left(1-\bar{x}_{\hi}\right)\Delta^{2}_{\delta {\rm x}}(k) \nonumber \\
& & + (1-\bar{x}_{\hi})^{2}\Delta^{2}_{\rm xx}(k),
\end{eqnarray}
where $\Delta^{2}_{\delta\delta}$, $\Delta^{2}_{\delta {\rm x}}$ and $\Delta^{2}_{\rm xx}$ are the matter density, density-ionization and
ionization PS respectively and fluctuations in the ionized fraction are defined as 
$\delta_{\rm x} = (\bar{x}_{\hii{}} - \langle \bar{x}_{\hii{}} \rangle)/\langle \bar{x}_{\hii{}} \rangle)$. Although \citet{Lidz:2007p1929} 
showed that this first-order approximation provides a rather poor description for the 21 cm PS, for our qualitative discussion 
below equation~(\ref{eq:decomp}) sufficiently highlights the major components to the total power.

Before reionization gets well underway ($\bar{x}_{\hi{}} > 0.9$) $\Delta^{2}_{21}$ closely 
follows the matter density PS, $\Delta^{2}_{\delta\delta}$. Initially, cosmic \hii{} bubbles grow around biased density 
peaks hosting galaxies.  This causes the large-scale power to drop due to the increased (negative) contribution from 
the cross-correlation term, $\Delta^{2}_{\delta {\rm x}}$. As time evolves ($\bar{x}_{\hi{}}$ decreases), the \hii{} bubbles 
become more numerous (with the increasing population of ionizing sources) and begin to overlap, increasing the contribution 
from the ionization PS, $\Delta^{2}_{\rm xx}$. Eventually, this term begins to dominate, resulting in an overall increase in the 
total power in $\Delta^{2}_{21}$ across all scales. At $\bar{x}_{\hi{}} = 0.5$, the large-scale power peaks. As the \hii{} bubbles 
grow and overlap, the peak power in $\Delta^{2}_{21}$ shifts to larger physical scales (smaller $k$) as $\Delta^{2}_{\rm xx}$
is sensitive to the physical size of the \hii{} regions. At the same time, the power on small scales begins to drop as the number of 
small, isolated \hii{} regions decreases.

This generic PS evolution during the EoR is evident in the first three panels of Fig.~\ref{fig:3param_variation} 
(see also, e.g. \citealt{McQuinn:2007p1665,Lidz:2008p1744,Friedrich:2011p1816}).  A larger value of $\zeta_{0}$ 
corresponds to a more advanced EoR at $z=9$, as galaxies are more efficient at producing ionizing photons.  
Similarly, a smaller value of $T^{\rm Feed}_{\rm vir}$ results in a more advanced EoR, as smaller, more abundant 
galaxies can efficiently form stars.  The progress of the EoR is less sensitive to variation in $R_{\rm mfp}$ (second panel); 
instead this maximum horizon for the ionizing photons produces a prominent `knee' in the PS, suppressing coherent 
ionization structure on large scales (e.g. \citealt{Alvarez:2012p1930,Mesinger:2012p1131}).

The PS is not however entirely determined by $\bar{x}_{\hi{}}$. In the final panel of Fig.~\ref{fig:3param_variation} we 
illustrate the impact of our \eor{} parameters at a fixed \igm{} neutral fraction of $\bar{x}_{\hi{}} = 0.71$ (i.e.\ all at the 
equivalent stage of reionization). For example, if the EoR is driven by more biased, larger mass haloes 
(larger $T^{\rm Feed}_{\rm vir}$), their relative dearth must be compensated for by a higher ionizing efficiency.  This results in an EoR 
morphology with larger, more isolated cosmic \hii{} patches \citep[e.g.][]{McQuinn:2007p1665}.  $\Delta^{2}_{\rm xx}$ is more 
peaked on large scales, with a decrease in $\Delta^{2}_{\delta {\rm x}}$.  As a result, the 21 cm PS is higher, with a more 
prominent large-scale `bump' tracing the typical \hii{} region size.  Also explicitly evident in the last panel is the prominent 
`knee' feature, imprinted by small values of $R_{\rm mfp}$. Here, for $R_{\rm mfp} = 3$~Mpc (blue curve), the ionization 
PS peaks on much smaller scales, with a sharp drop in large-scale power compared to the other curves computed with 
$R_{\rm mfp} = 15$~Mpc.  Therefore, the 21 cm PS contains information on {\it both} the EoR epoch (i.e. $\bar{x}_{\hi{}}$), 
as well as {\it independent} (i.e. at fixed $\bar{x}_{\hi{}}$) information on the properties of galaxies and the IGM.

In all panels, we provide the estimated 1$\sigma$ errors on the 21 cm PS for each of the considered 21 cm experiments, 
including the additional 25 per cent modelling uncertainty. It is immediately obvious that LOFAR will have difficulty discriminating 
among various models, having to rely on the largest scales close to the foreground-dominated regime. HERA and the SKA on 
the other hand should be able to recover meaningful constraints on \eor{} parameters, with the large-scales limited by the 
additional 25 per cent modelling uncertainty we include. On smaller spatial scales, SKA performs considerably better than HERA.

\subsection{Single epoch observation of the 21 cm PS} \label{sec:HERA_SKA}

21 cm experiments have coverage over a large bandwidth (e.g. 50--350~MHz for the SKA allowing for observations to 
$z\lesssim28$). However, foregrounds and high data rates can limit the coverage to narrower instantaneous bandwidths. 
Hence, we begin by analysing an observation at a single redshift (assuming our fiducial bandwidth of 8 MHz), before moving on to 
a broader bandpass observation. We restrict our analysis to only second-generation 21 cm experiments, HERA and the SKA, 
since with a single bandwidth the first-generation instruments will only achieve a marginal detection at best 
(e.g. \citealt{Mesinger:2014p244, Pober:2014p35}; see also Fig. \ref{fig:3param_variation}).

 \begin{figure*} 
	\begin{center}
		\includegraphics[trim = 0.2cm 0.5cm 0cm 0.5cm, scale = 0.85]{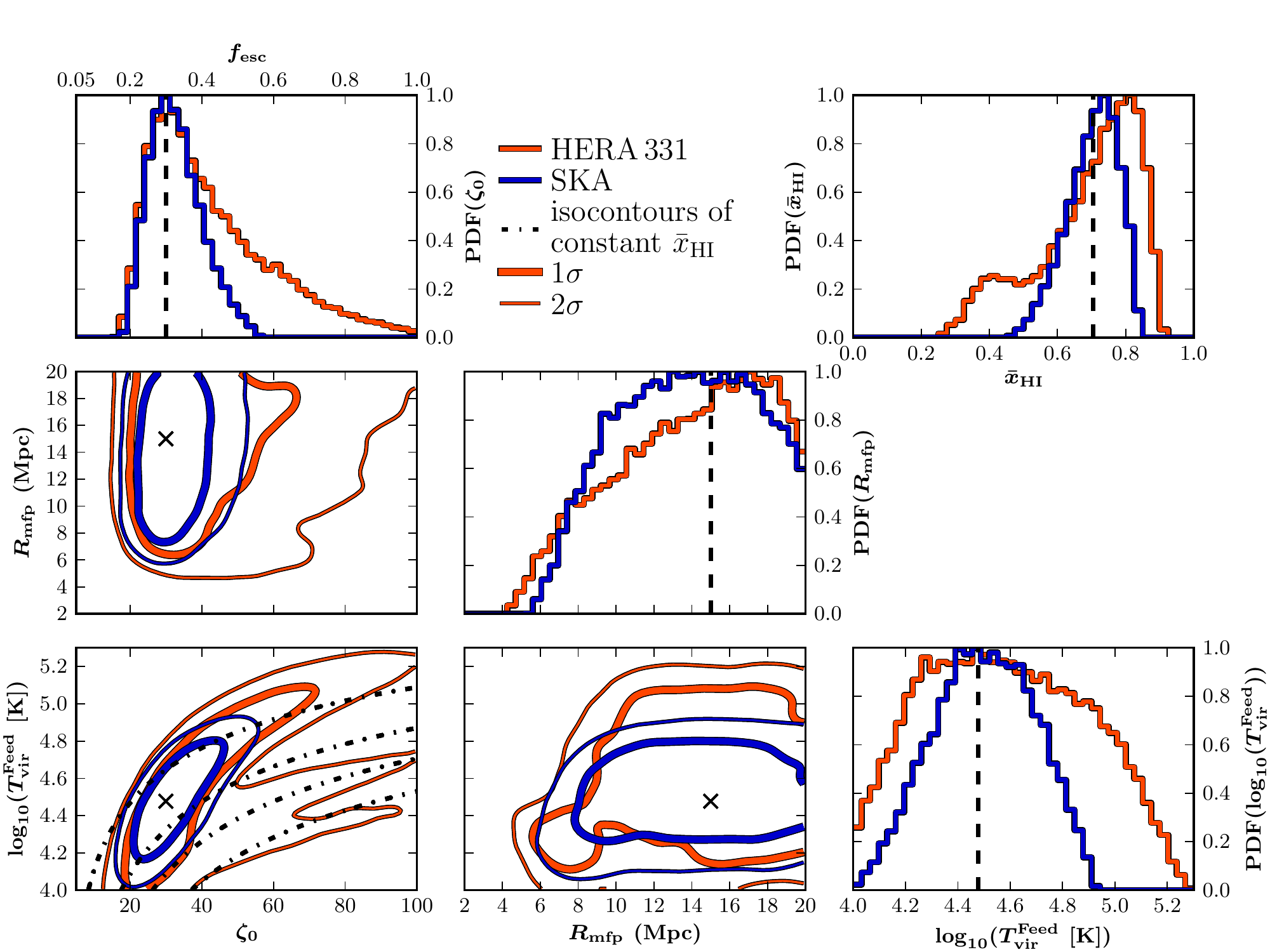}
	\end{center}
\caption[]
{The recovered constraints from \cmmc{} on our three parameter \eor{} model parameters for a single ($z=9$) 1000 h observation
 of the 21 cm PS obtained with HERA (red curve) and the SKA (blue curve). In the diagonal panels we provide the 1D 
 marginalized PDFs for each of our \eor{} model parameters 
 ($\zeta_{0}$, $R_{\rm mfp}$ and  log$_{10}$$(T^{\rm Feed}_{\rm vir})$ respectively) and we highlight our fiducial choice
 for each by the vertical dashed line. Additionally, we cast our ionizing efficiency, $\zeta_{0}$, into a corresponding escape 
 fraction, $f_{\rm esc}$, on the top axis (simply using the fiducial values in equation~\ref{eq:Zeta}). In the upper right panel we 
 provide the 1D PDF of the recovered \igm{} neutral fraction where the vertical dashed line corresponds to the neutral 
 fraction of the mock 21 cm PS observation ($\bar{x}_{\hi{}} = 0.71$). Finally, in the lower left corner we provide the 
 1 (thick) and 2$\sigma$ (thin) 2D joint marginalized likelihood contours for our three \eor{} parameters (crosses denote 
 their fiducial values, and the dot--dashed curves correspond to isocontours for $\bar{x}_{\hi{}}$ of 20, 40, 60 and 80 per cent 
 from bottom to top).}
\label{fig:2D_HERA_SKA_singlez}
\end{figure*}	

In Fig.~\ref{fig:2D_HERA_SKA_singlez}, we compare the outputs of \cmmc{} for both HERA (red curve) and 
 SKA (blue curve) for an assumed single $z=9$ observation of our fiducial 21 cm PS with a total integration time of 1000 h. 
 In this figure, and for the remainder of this work we assume uniform priors on all recovered parameters within their 
 allowed ranges outlined in Section~\ref{sec:21CMFAST}.
 The dashed vertical lines denote the mock observation values. Across the diagonal panels of this figure we provide the 
 1D marginalized PDFs for each of our \eor{} parameters. In the top-right panel, we provide the marginalized 1D PDFs of 
 the \igm{} neutral fraction. Additionally, we choose to renormalize all 1D PDFs to have the peak probability equal to 
 unity to better visually emphasize the shape and width of the recovered distribution. In the lower-left corner of 
 Fig.~\ref{fig:2D_HERA_SKA_singlez}, we show the 2D joint marginalized likelihood distributions. In each panel, the 
 cross denotes the location of our fiducial \eor{} parameters. For each, we construct both the 1$\sigma$ and 2$\sigma$ contours 
 (thick and thin lines, respectively), by computing a smoothed 2D histogram of the entire MCMC sample and estimating 
 the likelihood contours which enclose 68 and 95 per cent of the data sample, respectively.
 
For all \eor{} parameters the recovered PDFs are centred around their input fiducial parameters, 
 highlighting the strong performance of \cmmc{} at recovering our input model. This is despite the 
 asymmetric distributions of the recovered PDFs, emphasizing the strength of the MCMC approach within \cmmc. 
 Comparing these two 21 cm experiments, we note the significant improvement achievable with the increased 
 sensitivity of the SKA. This can easily be seen in the case of both $\zeta_{0}$ and $T^{\rm Feed}_{\rm vir}$,
 where the widths of the 1D PDFs, as well as the 2D joint likelihood contours, are noticeably tighter for SKA than for HERA.
  
The improved constraints from the SKA relative to HERA can be best explained by referring to Fig.~\ref{fig:3param_variation}. 
 We observe that while the sensitivity at large scales for both SKA and HERA are comparable, only the SKA is capable of 
 probing the available small-scale information. Given the degenerate nature of the \eor{} parameters, by accessing this 
 additional shape information tighter constraints are achievable. While it is not sufficient to break the degeneracies it does 
 reduce their amplitude, substantially limiting the allowed \eor{} parameter space. It is important to note that with only a 
 single redshift observation, the uncertainty is large and the likelihood is non-Gaussian distributed. Nevertheless, our 
 Bayesian MCMC framework captures parameter constraints and degeneracies, recovering the actual shape of the 
 likelihood distribution (without the Gaussian assumptions of the Fisher matrix formalism; \citealt{Pober:2014p35}).
  
  \begin{table*}
\begin{tabular}{@{}lccccccc}
\hline
Instrument & & Parameter & & & $\bar{x}_{\hi{}}$ &  \\
(single-$z$) & $\zeta_{0}$ & $R_{\rm mfp}$ & log$_{10}$$(T^{\rm Feed}_{\rm vir})$ & & $z = 9$ & & \\
\hline
\vspace{0.8mm}
HERA 331 & $38.38\substack{+22.43 \\ -11.33}$ & $14.34\substack{+3.72 \\ -4.95}$ & $4.57\substack{+0.36 \\ -0.32}$ & & $0.73\substack{+0.10 \\ -0.21}$ & \\
\vspace{0.8mm}
HERA 331 (with prior) & $38.37\substack{+20.86 \\ -11.53}$ & $13.88\substack{+3.76 \\ -4.39}$ & $4.65\substack{+0.32 \\ -0.35}$ & & $0.75\substack{+0.09 \\ -0.13}$ & \\
\vspace{0.8mm}
SKA & $32.11\substack{+8.32 \\ -6.47}$  & $13.78\substack{+3.87 \\ -4.09}$ & $4.50\substack{+0.20 \\ -0.21}$ & & $0.71\substack{+0.07 \\ -0.09}$& \\
\hline
\end{tabular}
\caption{The median recovered values (and associated 16th and 84th percentile errors) for our three \eor{} 
model parameters, $\zeta_{0}$, $R_{\rm mfp}$ and $T^{\rm Feed}_{\rm vir}$ and the associated \igm{} neutral fraction, 
$\bar{x}_{\hi{}}$. We assume a total 1000 h integration time for a single epoch ($z=9$) observation of the 21 cm 
PS and compare the recovered constraints for HERA, with and without a $z=7$ \igm{} neutral fraction prior ($\bar{x}_{\hi{}} > 0.05$),
 and for the SKA 
respectively. Our fiducial parameter set is 
($\zeta_{0}$, $R_{\rm mfp}$, ${\rm log}_{10}T^{\rm Feed}_{\rm vir}$) = (30, 15~Mpc, 4.48) which
results in an \igm{} neutral fraction of $\bar{x}_{\hi{}} = 0.71$.}
\label{tab:3param_summary_singlez}
\end{table*}   
  
In the $\zeta_{0}$-${\rm log}_{10}T^{\rm Feed}_{\rm vir}$ plane for HERA in lower-left panel of 
Fig.~\ref{fig:2D_HERA_SKA_singlez}, we observe at the 95 per cent confidence level a multimodal distribution for 
$\zeta_{0}$ and $T^{\rm Feed}_{\rm vir}$ as highlighted by the lower streaks for high $\zeta_{0}$ and low $T^{\rm Feed}_{\rm vir}$.
This region of parameter space is capable of reionizing the \igm{} earlier than our fiducial \eor{} model, due to a 
brighter population of lower mass galaxies (see Section~\ref{sec:variation}). This is highlighted by overlaying the 
$\bar{x}_{\hi{}}$ isocontours for the \eor{} parameter space, where it is clear that these two streaks reproduce different
\igm{} neutral fractions. Correspondingly, these models result in a smaller, secondary feature around $\bar{x}_{\hi{}} = 0.4$ in the 
\igm{} neutral fraction PDF. Models in this region of the \eor{} parameter space have similar large-scale 21 cm power, but 
different small-scale structure which HERA cannot differentiate. The SKA, due to its higher sensitivity on small scales does 
not exhibit the same behaviour.  Alternatively, these degeneracies can be broken with additional redshift observations which 
constrain the redshift evolution of the large-scale power (see below).
  
In order to quantitatively assess the performance of \cmmc{} we choose to report the median value of each \eor{} 
parameter and the associated 16th and 84th percentiles. This choice follows on from the fact that the recovered 
1D PDFs do not need to be normally distributed within the Bayesian MCMC framework, as observed in the case of HERA in 
Fig.~\ref{fig:2D_HERA_SKA_singlez}. In Table~\ref{tab:3param_summary_singlez} we summarize the recovered \eor{}
parameter constraints for each of the 21 cm experiments from a single epoch observation of the 21 cm PS. 

For HERA, we note that while both $R_{\rm mfp}$ and 
 $T^{\rm Feed}_{\rm vir}$ (also $\bar{x}_{\hi{}}$), are recovered to within 5 per cent of our fiducial parameters, in the case
 of $\zeta_{0}$ we over predict the expected value by close to 30 per cent. This is not surprising given the 
 mild positive asymmetry observed in the recovered 1D PDF for $\zeta_{0}$. Since we choose to report the median 
 recovered value for all \eor{} parameters, whenever we recover an asymmetric distribution our 
 reported value can be shifted away from the input fiducial value. Despite the skewness, the 16th and 
 84th percentiles should for the most part enclose our fiducial values.

In the case of the SKA, we are able to recover all \eor{} parameters values at 
 less than 10 per cent deviation from their fiducial values. While the errors on $R_{\rm mfp}$ remain
 relatively consistent with those for HERA, we observe a reduction of close to a factor of
 2 (3) for the lower (upper) 1$\sigma$ bounds on 
 $\zeta_{0}$ and a 30 per cent reduction on the 1$\sigma$ errors on $T^{\rm Feed}_{\rm vir}$. We note a reduction also of 
 approximately 25 per cent on the recovered 1$\sigma$ errors for the \igm{} neutral fraction.

\begin{figure*} 
	\begin{center}
		\includegraphics[trim = 0.2cm 0.5cm 0cm 0.5cm, scale = 0.85]{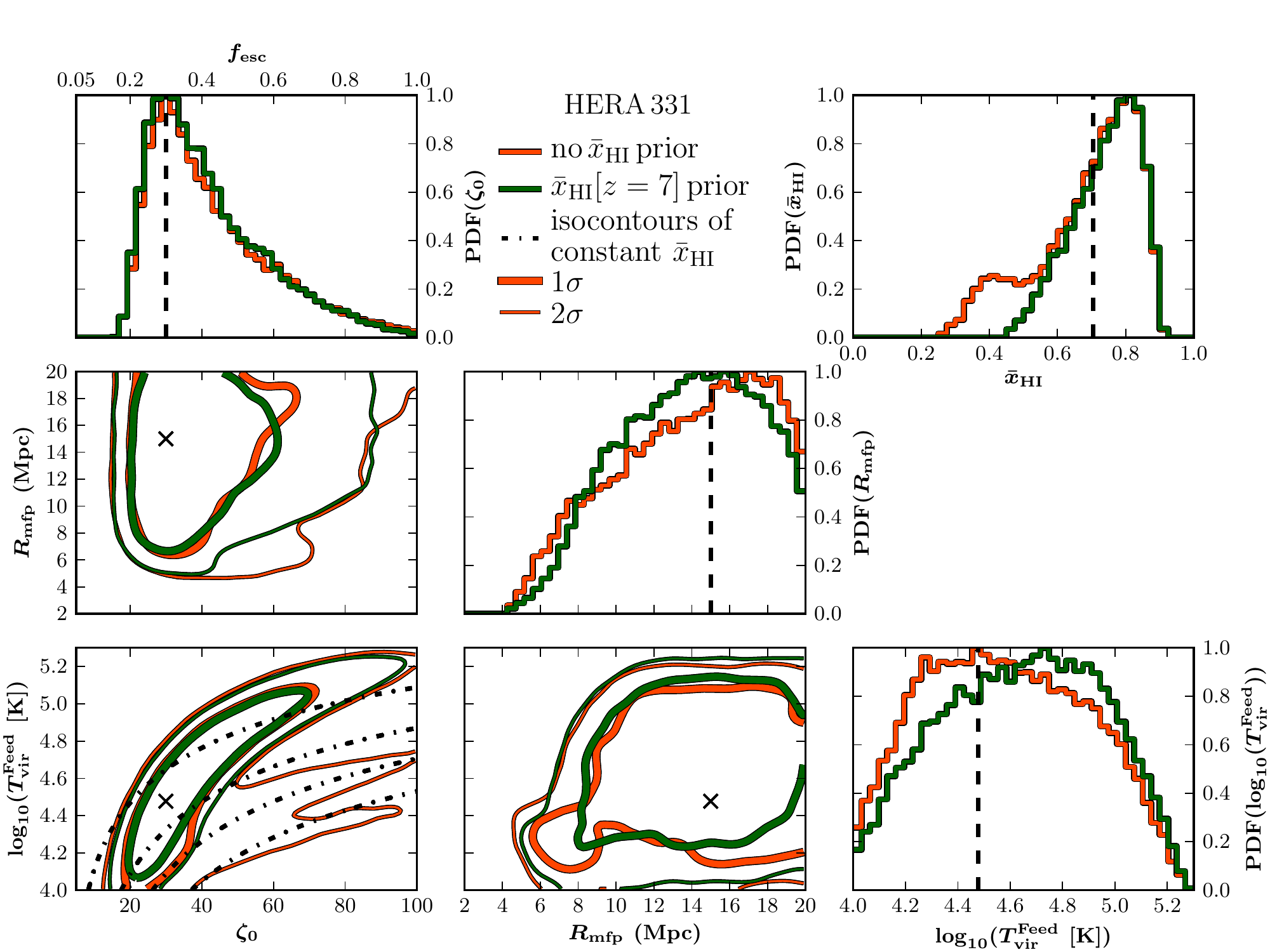}
	\end{center}
\caption[]
{Similar to Fig.~\ref{fig:2D_HERA_SKA_singlez}, except now we compare the impact of including an \igm{} neutral fraction 
prior for a single ($z=9$) 1000 h observation with HERA. We consider an \igm{} neutral fraction prior at $z=7$ of 
$\bar{x}_{\hi{}} > 0.05$ (a conservative choice motivated by recent QSO and LAE observations; green curves) 
and without a neutral fraction prior (red curve).
The inclusion of a neutral fraction prior improves the constraints on the reionization parameters from a single $z$ observation.}
\label{fig:2D_HERA_prior}
\end{figure*}

\subsection{Single redshift observation with a neutral fraction prior} \label{sec:NFPrior}

In the previous section, we noted that the same large-scale power could occur at different \igm{} neutral fractions. This 
degeneracy can be broken with increased small-scale sensitivity (as in the case of SKA).  Here, we investigate whether 
we can improve upon the available constraining power from a single redshift observation by instead including a prior 
on the \igm{} neutral fraction from observations at the tail end of the reionization epoch. For example, such constraints
can be obtained from the spectra of high-$z$ quasars such as ULAS J1120+0641 \citep{Bolton:2011p1063,Mortlock:2011p1049} 
or from the observed drop in Lyman-$\alpha$ emission from high-$z$ galaxies 
\citep[e.g.][]{Caruana:2014p1585,Pentericci:2014p1531,Dijkstra:2014p3264,Mesinger:2015p1584}. 
Here, we consider a purely illustrative conservative lower limit on the \igm{} neutral fraction at $z=7$ of 5 per cent. 
We caution however that priors on different epochs can make our results more model dependent. For example, our 
fiducial \eor{} model does not allow for a redshift evolution in the reionization parameters. This is straightforward to 
account for, however, we postpone it to future work.

In Fig.~\ref{fig:2D_HERA_prior}, we compare the recovered \eor{} parameter constraints for HERA from the same 1000 h 
observation of the 21 cm PS at $z=9$, following the inclusion of the \igm{} neutral fraction prior at $z=7$. Additionally, in 
Table~\ref{tab:3param_summary_singlez}, we again provide the recovered median and 1$\sigma$ errors for our \eor{} parameters. 
After the inclusion of the $z=7$ neutral fraction prior we observe only a very marginal improvement in either the median 
recovered values or their 1$\sigma$ errors and this is equivalently shown by the slightly narrower widths of the recovered 1D PDFs. 

However, in the case of the 2D joint likelihood distributions we are able to observe an improvement in the allowed \eor{} parameter 
space. Specifically, focusing again on the $\zeta_{0}$-${\rm log}_{10}T^{\rm Feed}_{\rm vir}$ plane, while the 1$\sigma$ errors 
are only marginally improved, the 2$\sigma$ contours are notably reduced. Most importantly, the \igm{} neutral fraction 
prior removes the second degenerate streak which marginally allowed a significantly different reionization history in the absence of the
prior when observed with HERA. This can clearly be seen from the removal of the previously noted secondary bump at  
$\bar{x}_{\hi{}} = 0.4$ in the \igm{} neutral fraction PDF. By imposing that the \igm{} must at least be 5 per cent neutral at $z=7$, 
this rules out reionization scenarios which would otherwise predict an early end to the reionization epoch from a single 
redshift observation of the 21 cm PS. This highlights the importance of the additional leverage gained from probing the reionization 
history at improving the \eor{} reionization constraints, especially in the absence of strong telescope sensitivity at small scales.

The level of improvement achievable from an \igm{} neutral fraction prior on a single epoch measurement of the 21 cm PS depends on 
numerous factors. First, imposing a harder prior than our conservative choice would allow increased constraining power
on the allowed \eor{} parameter space. Secondly, with decreasing telescope sensitivity, a neutral fraction 
prior becomes more important, as knowledge of the EoR history compensates for poor single epoch PS coverage.
It is feasible that with improved foreground removal strategies, a marginal detection of the 21 cm PS could be
achieved with first generation instruments \citep{Pober:2014p35}. Therefore, in this scenario, a neutral fraction prior would be of 
vital importance for improving the available \eor{} constraining power.

Finally, due to the flexibility of \cmmc{} we emphasize that it is straightforward to additionally include direct priors on our 
model \eor{} parameters motivated by current and upcoming observations and theoretical advances. 
Given the observed degeneracies between the \eor{} parameters within this model, it is clear that
including priors on our model parameters would notably improve the overall constraining power.

\begin{table*}
\begin{tabular}{@{}lccccccc}
\hline
Instrument & & Parameter & & & $\bar{x}_{\hi{}}$ &  \\
(multiz-$z$) & $\zeta_{0}$ & $R_{\rm mfp}$~(Mpc) & log$_{10}$$(T^{\rm Feed}_{\rm vir})$ & $z = 8$ & $z = 9$ & $z = 10$\\
\hline
\vspace{0.8mm}
LOFAR & $45.21\substack{+26.27 \\ -18.06}$ & $12.98\substack{+ 4.44\\ -5.50}$ & $4.73\substack{+0.34 \\ -0.32}$ & $0.57\substack{+ 0.20\\ -0.21}$ & $0.76\substack{+0.13 \\ -0.15}$ & $0.88\substack{+ 0.07\\ -0.10}$\\
\vspace{0.8mm}
HERA 331 & $30.17\substack{+7.60 \\ -5.35}$ & $15.13\substack{+ 2.92\\ -3.01}$ & $4.49\substack{+ 0.15\\ -0.16}$ & $0.49\substack{+ 0.05\\ -0.05}$ & $0.71\substack{+ 0.04\\ -0.04}$ & $0.84\substack{+ 0.03\\ - 0.03}$ \\
\vspace{0.8mm}
SKA & $30.04\substack{+ 5.65\\ -4.48}$ & $15.22\substack{+ 2.82 \\ -2.91}$ & $4.48\substack{+ 0.11\\ - 0.12}$ & $0.49\substack{+0.04 \\ -0.04}$ & $0.71\substack{+ 0.03\\ -0.03}$ & $0.84\substack{+ 0.02\\ - 0.02}$ \\
\hline
\vspace{0.8mm}
HERA 331 (w/o 25 per cent modelling uncertainty) & $28.78\substack{+ 2.54\\ -1.98}$ & $14.33\substack{+ 1.12 \\ -0.93}$ & $4.47\substack{+ 0.06\\ - 0.06}$ & $0.47\substack{+0.02 \\ -0.02}$ & $0.69\substack{+ 0.02\\ -0.02}$ & $0.82\substack{+ 0.01\\ - 0.01}$ \\
\vspace{0.8mm}
SKA (w/o 25 per cent modelling uncertainty) & $30.31\substack{+ 1.70\\ -1.62}$ & $15.47\substack{+ 1.41 \\ -1.20}$ & $4.48\substack{+ 0.04\\ - 0.04}$ & $0.48\substack{+0.01 \\ -0.01}$ & $0.71\substack{+ 0.01\\ -0.01}$ & $0.83\substack{+ 0.01\\ - 0.01}$ \\
\hline
\end{tabular}
\caption{Summary of the median recovered values (and associated 16th and 84th percentile errors) for our three \eor{} 
model parameters, $\zeta_{0}$, $R_{\rm mfp}$ and $T^{\rm Feed}_{\rm vir}$ and the associated \igm{} neutral fraction, 
$\bar{x}_{\hi{}}$. We assume a total 1000 h integration time for each 21 cm PS observation at $z=8$, 9 and 10 
for LOFAR, HERA and the SKA, respectively. For both HERA and the SKA, we provide the recovered median 
values both including and excluding our 25 per cent modelling uncertainty (see Section~\ref{sec:ModUncert}). 
Our fiducial parameter set is 
($\zeta_{0}$, $R_{\rm mfp}$, ${\rm log}_{10}T^{\rm Feed}_{\rm vir}$) = (30, 15~Mpc, 4.48) which
results in an \igm{} neutral fraction of $\bar{x}_{\hi{}} = 0.48$, 0.71, 0.83 at $z=8$, 9 and 10 respectively.}
\label{tab:3param_summary}
\end{table*}   

\subsection{Combining multiple redshift observations} \label{sec:multiz}  

Here, we consider combining multiple epoch observations of the 21 cm PS, assuming a broader instantaneous bandwidth is
available for the \eor{} observations. However, this comes at the cost of increased model dependence. For example, 
our three empirical \eor{} parameters are redshift-independent, whereas physically, we would expect each parameter to vary with 
redshift across the reionization epoch. In this sense, our astrophysical parameters can be thought of as `effective' EoR properties, 
averaging over any redshift evolution. In future, we will explicitly include a generalized redshift evolution for \eor{} parameters 
(e.g. \citealt{Barkana:2009p116, Mesinger:2012p1131,Zahn:2012p1156}).

 \begin{figure*} 
	\begin{center}
		\includegraphics[trim = 0.2cm 0.5cm 0cm 0.5cm, scale = 0.85]{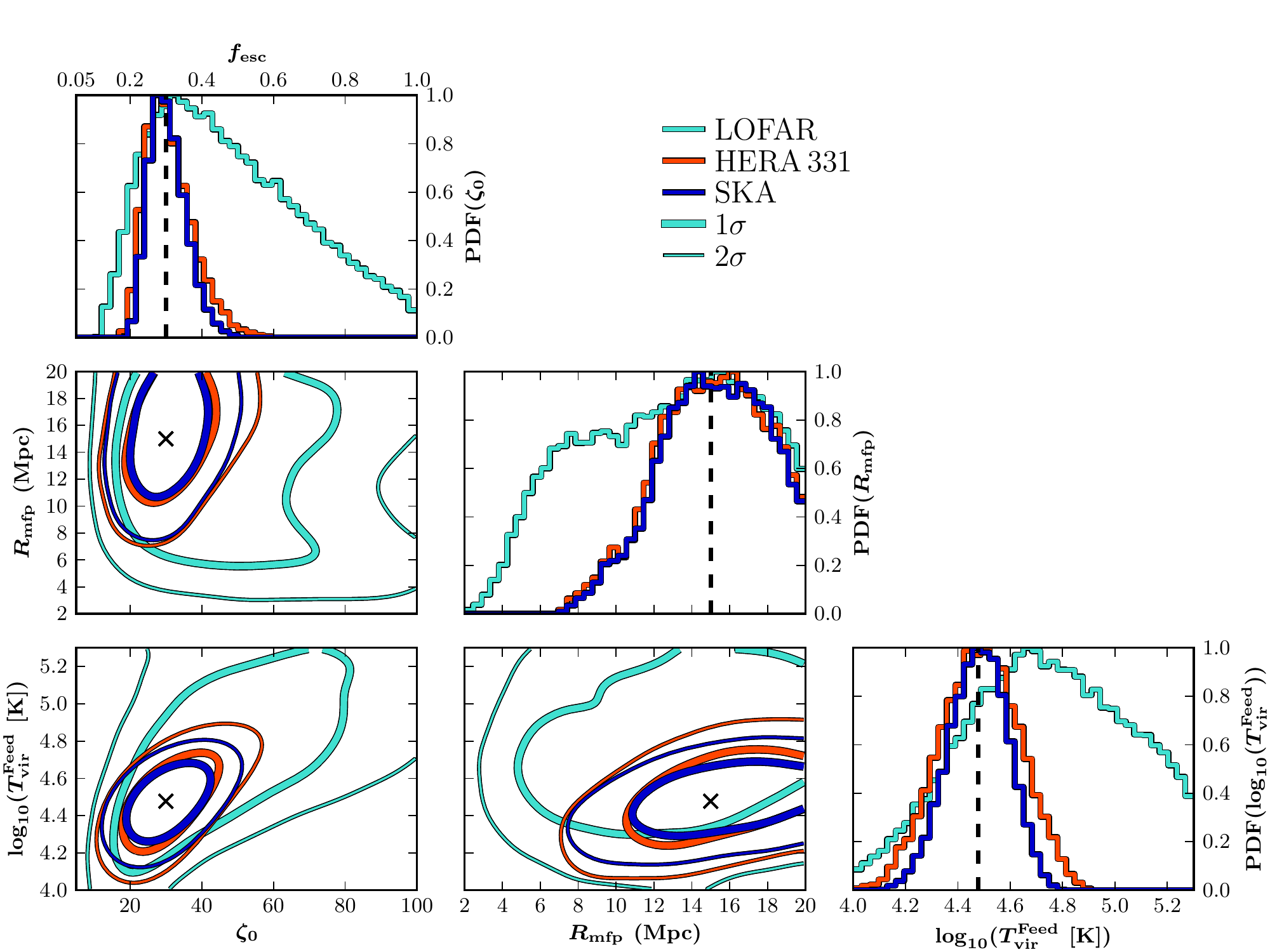}
	\end{center}
\caption[]
{The recovered constraints from \cmmc{} on our reionization model parameters from combining three independent
 ($z=8, 9$~and~10) 1000 h observations of the 21 cm PS. We compare the different telescope arrays, LOFAR (turquoise), 
 HERA (red) and SKA (blue). Across the diagonal panels we provide the 1D marginalized PDFs for each of 
 our reionization parameters [$\zeta_{0}$, $R_{\rm mfp}$ and  log$_{10}$$(T^{\rm Feed}_{\rm vir})$ respectively] and 
 highlight our fiducial model parameter by a vertical dashed line. Additionally, we provide the 1 (thick) and 2$\sigma$ (thin) 
 2D joint marginalized likelihood contours for our three reionization parameters in the lower left corner of the 
 figure (crosses mark our fiducial reionization parameters). Including redshift information can significantly improve the 
 constraints on each reionization parameter, and could even yield sufficient information to allow LOFAR to 
 provide marginal constraints.}
\label{fig:2D_LOFAR_HERA_SKA_multiz}
\end{figure*}

\begin{figure*} 
	\begin{center}
		\includegraphics[trim = 0.2cm 0.5cm 0cm 0.45cm, scale = 0.9]{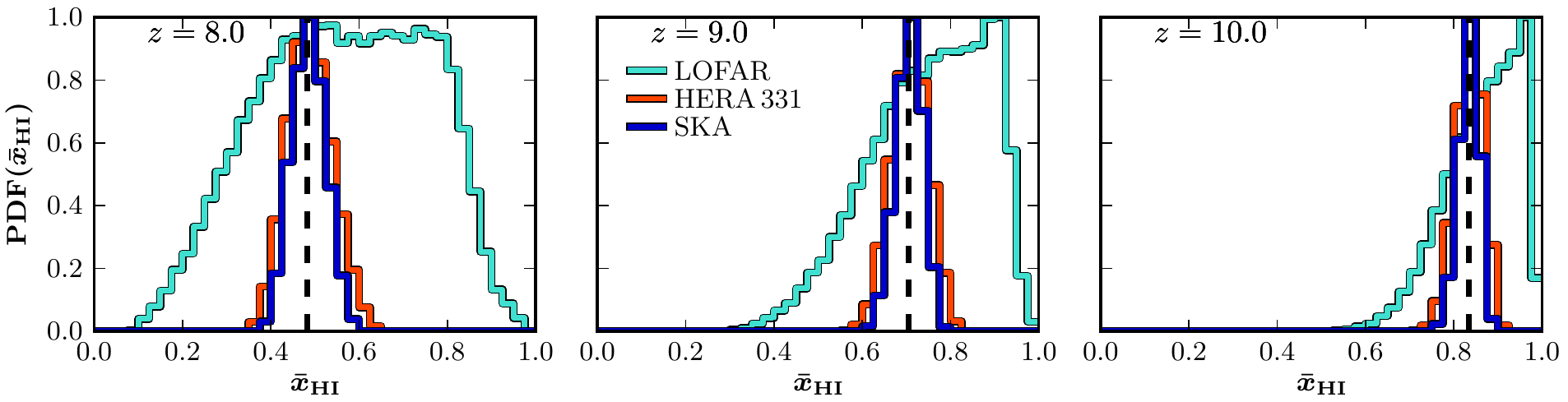}
	\end{center}
\caption[]
{1D PDFs of the recovered \igm{} neutral fraction from our 1000 h observations of the 21 cm PS at $z=8$ (left), 
$z=9$ (centre) and $z=10$ (right). Colours are the same as Fig.~\ref{fig:2D_LOFAR_HERA_SKA_multiz} and vertical 
dashed lines denote the neutral fraction of the mock 21 cm PS observations, $\bar{x}_{\hi{}}$ = 0.48, 0.71 and 0.83 respectively.}
\label{fig:LOFAR_HERA_SKA_multiz_NF}
\end{figure*}

To combine all the available information across the multiple epochs and differentiate between the sampled \eor{} parameters, 
we compute the total maximum likelihood fit of the \eor{} model, obtained by linearly combining the individual $\chi^{2}$ statistics 
from each redshift. In principle, we could instead perform a weighted summation to estimate the total maximum likelihood; however,
 in neglecting to do so our errors are more conservative. 
  
In Fig.~\ref{fig:2D_LOFAR_HERA_SKA_multiz}, we compare the recovered 1D marginalized PDFs and 
2D marginalized likelihood contours for our three fiducial \eor{} parameters ($\zeta_{0}$, $R_{\rm mfp}$ and 
log$_{10}$$(T^{\rm Feed}_{\rm vir})$). For this, we assume three independent 1000 h observations 
of the 21 cm PS at $z=8$, 9 and 10 for LOFAR (turquoise), HERA (red) and SKA (blue) each with a 1000 h integration time. In 
Fig.~\ref{fig:LOFAR_HERA_SKA_multiz_NF} we provide the recovered 1D PDFs of the \igm{} neutral fraction 
from each individual epoch. Note, for our fiducial \eor{} model, the corresponding \igm{} neutral fractions are
$\bar{x}_{\hi{}}$ = 0.48, 0.71 and 0.83, respectively. Finally, in Table~\ref{tab:3param_summary} we provide the 
recovered median, 16th and 84th percentiles for each \eor{} parameter and the corresponding \igm{} neutral fraction.
 
From Fig.~\ref{fig:2D_LOFAR_HERA_SKA_multiz}, we observe the 1D PDFs and joint 2D likelihood contours
 to be almost indistinguishable between HERA and the SKA, with the SKA performing only marginally better. Furthermore, 
 comparing with Fig.~\ref{fig:2D_HERA_SKA_singlez}, the widths of both the 1D PDFs and 2D likelihood contours 
 have significantly reduced, drastically in the case of HERA. This emphasizes the strength of combining multiple epoch
 observations of the 21 cm PS (under the simple assumption they can be linearly combined). Quantitatively, we find 
 little difference between the recovered constraints, with both instruments recovering median values of our \eor{} 
 parameters to better than 1.5 per cent. Due to the improved constraining power from sampling the reionization history
 the recovered PDFs are now normally distributed. Therefore, we can estimate the total 1$\sigma$ fractional error for each 
 parameter. For SKA (HERA), we find errors of 16.7 (22.0) on $\zeta_{0}$, 17.8 (18.4) on $R_{\rm mfp}$ 
 and 2.4 (3.3) per cent on log$_{10}$$(T^{\rm Feed}_{\rm vir})$. Additionally, across all three observed epochs, 
 both HERA and SKA tightly recover the expected \igm{} neutral fraction of our fiducial \eor{} model 
 (see Fig.~\ref{fig:LOFAR_HERA_SKA_multiz_NF}).
 
Importantly, while for a single epoch observation, the SKA notably outperforms HERA, the improvements available when 
 probing the reionization history can make these two instruments comparable. By combining multiple epoch observations, these 
 two experiments are able to pick up the same evolution in the 21 cm PS as the power rises and falls during reionization 
 (see, e.g. \citealt{Lidz:2008p1744} and our Section~\ref{sec:variation}), compensating for their different sensitivity at small scales 
 (for the case of these EoR models). Therefore, while the additional small-scale power can aid significantly in improving the 
 constraining power from an individual observation, it only marginally improves the constraining power across multiple epochs.
 
For LOFAR, we find that by jointly combining multiple epoch observations we are now capable of recovering reionization 
constraints. As expected though, the quality of these constraints is drastically reduced compared to the 
second-generation experiments. Additionally, we recover broad marginalized 1D PDFs, which encompass the majority
 of the \eor{} parameter space and their asymmetric nature complicates a quantitative comparison with the second-generation 
 experiments. We find LOFAR can only marginally rule out (at $\sim 2\sigma$) either a
 rapid reionization ($\bar{x}_{\hi{}} < 0.2$ at $z=8$) or a delayed/extended reionization ($\bar{x}_{\hi{}} > 0.9$ at $z=8$).
 However, this is still encouraging for the first generation of 21 cm experiments given our conservative
  foreground and sensitivity assumptions.

\begin{figure*} 
	\begin{center}
		\includegraphics[trim = 0.2cm 0.5cm 0cm 0.5cm, scale = 0.85]{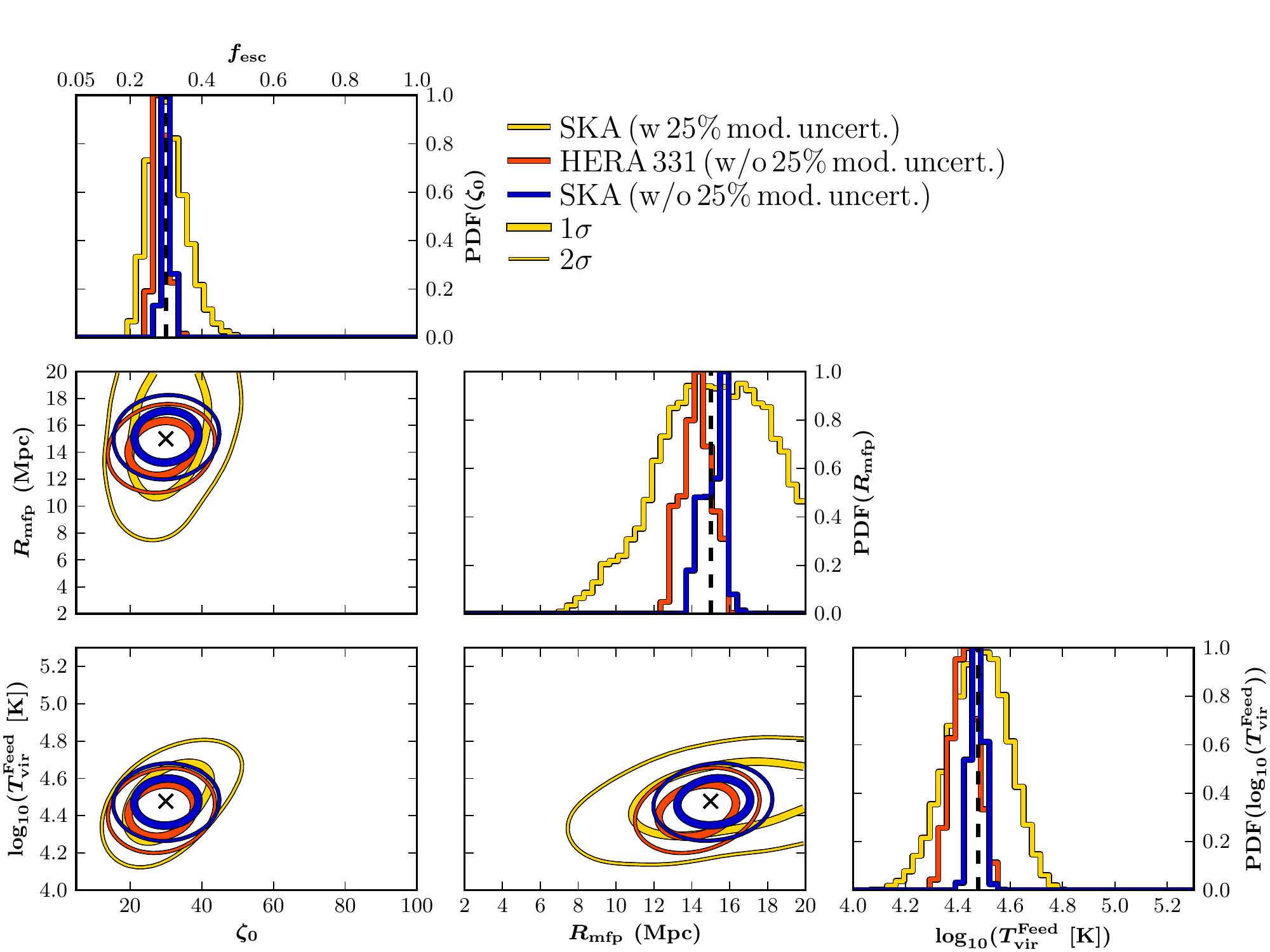}
	\end{center}
\caption[]
{Analogous to Fig.~\ref{fig:2D_LOFAR_HERA_SKA_multiz}, except here we show the impact of our assumed choice 
of a 25 per cent \eor{} modelling error, which dominates on large-scales (see Fig.~\ref{fig:3param_variation}). 
Constraints from the SKA with the modelling error are shown in yellow, while those without any modelling error are shown in blue (SKA) and red (HERA).
We see that under the optimistic 
assumption that we can perfectly characterize the \eor{} modelling uncertainties, the derived parameter constraints can be
improved by a factor of a few (see Table~\ref{tab:3param_summary}).}
\label{fig:2D_HERA_SKA_multiz_ModUncert}
\end{figure*}

\subsection{Impact of the \eor{} modelling uncertainty} \label{sec:ModUncert}

Thus far we have discussed the available constraints on our \eor{} parameters including a 25 per cent 
 \eor{} modelling uncertainty, which dominates on large scales (see Fig.~\ref{fig:3param_variation}). Characterization of these uncertainties (e.g. by calibrating to more detailed RT simulations) will be essential to maximizing the achievable scientific gains from upcoming second-generation interferometers.
 To emphasize this, in Fig.~\ref{fig:2D_HERA_SKA_multiz_ModUncert} we perform the same analysis as in the previous section,
  instead now removing the 25 per cent error for HERA and the SKA, corresponding to a perfect 
  characterization of the \eor{} modelling uncertainties.
  We also list the corresponding constraints in Table~\ref{tab:3param_summary}.
    As expected, the resulting fractional errors are dramatically improved by a 
  factor of 2--3 to 8.1/5.5, 7.2/8.1 and 1.4/0.9  per cent for HERA/SKA 
  (for $\zeta_{0}$, $R_{\rm mfp}$ and  log $T^{\rm Feed}_{\rm vir}$, respectively).

  The fact that the parameter constraints improve dramatically and in a comparable manner for both HERA and the SKA shows that the redshift evolution of large-scale power dominates the constraining power for this EoR model.  Unlike for narrow-band, single-$z$ observations in Fig.~\ref{fig:2D_HERA_SKA_singlez}, the additional leverage provided by small scales does not dramatically improve the parameter constraints if the redshift evolution of the large-scale power can be constrained.  We caution however that this conclusion is dependent on the EoR model.

  We also note that these values without the modelling error are in reasonable 
 agreement with the roughly comparable\footnote{A like-to-like comparison is difficult for several reasons. First, these authors 
 jointly combine four different epoch observations of the 21 cm PS compared to our three. Since instrument sensitivity increases 
 with decreasing redshift, improved constraining power is available from their additional $z=7$ observation (although it is not a 
 significant improvement as three redshift measurements will have already strongly constrained the reionization history). 
 Secondly, we consider only a 331 telescope antenna for HERA, unlike their 547 telescope design, while for the SKA 
 there is the previously noted numerical error in their total sensitivities.} constraints recovered by \citet{Pober:2014p35}, 
 suggesting the Fisher matrix approach provides a reasonably accurate description of available constraints for this simple EoR 
 parametrization.

 \section{Two galaxy populations with mass-dependent ionizing efficiencies}
 \label{sec:DPL}

In the previous section, we highlighted the strength of \cmmc{} at providing constraints on our model parameters from a 
simple single ionizing source population. However, physically we expect galaxy formation to be more complicated. 
Therefore, in this section, we consider a more flexible \eor{} model which allows for both a halo mass-dependent ionizing 
efficiency and multiple ionizing source populations.

\begin{figure} 
	\begin{center}
		\includegraphics[trim = 0.5cm 1cm 0cm 0cm, scale = 0.46]{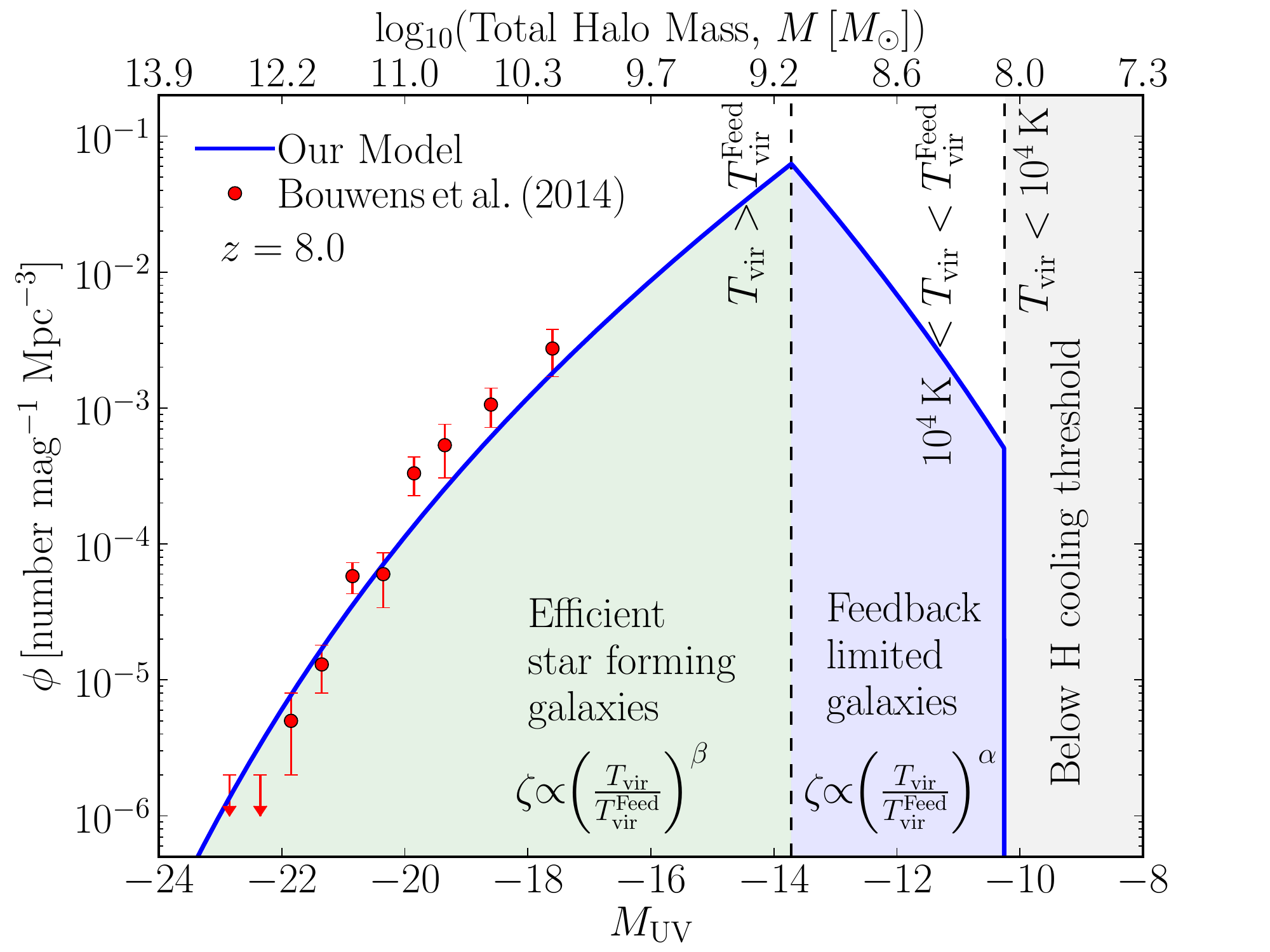}
	\end{center}
\caption[]
{A schematic picture of the $z=8$ non-ionizing UV luminosity function (blue curve) from our fiducial reionization model generalized 
to include two different source galaxy populations. First, above a critical temperature, $T^{\rm Feed}_{\rm vir}$, haloes are 
sufficiently massive to be weakly affected by internal feedback process (e.g.\ supernovae driven winds), whose efficiency 
we describe by the power-law index $\beta$. Between $T^{\rm Feed}_{\rm vir}$ and a cooling threshold temperature of 
$T_{\rm vir}=10^{4}\,{\rm K}$, haloes are sufficiently large to cool and form galaxies; however, their ionizing efficiency is affected by 
internal feedback processes, described by a second power-law index $\alpha$. Note that the model curve
at $T_{\rm vir} > T^{\rm Feed}_{\rm vir}$ is obtained from a power-law fit to the halo mass as a function of UV 
magnitude recovered from abundance matching, which we can use to imply} that for our fiducial choice of 
$\beta=-1/3$ we roughly obtain the scaling, $f_{\rm esc}\propto M^{-0.18}$ 
(see the text for further discussions).
\label{fig:UVLum_DPL}
\end{figure} 

\subsection{Generalized five parameter \eor{} model}

Rather than assuming a step-function for the ionizing efficiency as in the previous section, here we allow it to vary with the 
host halo potential both for the bright, star-forming galaxies and the faint, feedback limited galaxies,
\begin{eqnarray} \label{eq:zeta_M}
\zeta(T_{\rm vir}) = \zeta_{0}\left\{ \begin{array}{cl} 
      \left(\frac{T_{\rm vir}}{T^{\rm Feed}_{\rm vir}}\right)^{\beta}  & {T_{\rm vir} \geq T^{\rm Feed}_{\rm vir}}\\
      \left(\frac{T_{\rm vir}}{T^{\rm Feed}_{\rm vir}}\right)^{\alpha} & {10^{4}\,{\rm K} \leq T_{\rm vir} < T^{\rm Feed}_{\rm vir}}\\ 
      0 & {T_{\rm vir} < 10^{4}\,{\rm K}}.
\end{array} \right.
\end{eqnarray}
Here, $\zeta_{0}$, is simply the normalization of the ionization efficiency and retains the same definition as in Section~\ref{sec:Zeta}. 
When $\beta = 0$ and $\alpha\rightarrow \infty$, this expression is equivalent to the step function ionizing efficiency we considered 
earlier. A mass-dependent ionizing efficiency has previously been investigated to describe the ionizing galaxy population; however, 
these only considered a single population of galaxies described by a single power-law index 
\citep[e.g.][]{Furlanetto:2006p1647,McQuinn:2007p1665,Paranjape:2014p121}. A sharp mass-scale separating feedback limited 
and non-limited galaxies is expected in some theoretical models \citep[e.g.][]{Finlator:2011p3408,Raicevic:2011p3440}.
 
Our generalized \eor{} model contains a total of five parameters, the same three \eor{} parameters as described in 
Section~\ref{sec:21CMFAST} ($\zeta_{0}$, $R_{\rm mfp}$ and $T^{\rm Feed}_{\rm vir}$) and the two new power-law indices 
describing the mass-dependent ionizing efficiency, $\alpha$ and $\beta$. For our fiducial mock observation, we choose to adopt 
 $\alpha = 4/3$ and $\beta = -1/3$, both within the allowed range, $\alpha$, $\beta\in[-4/3,10/3]$. If instead we define 
 equation~(\ref{eq:zeta_M}) with respect to mass, these power-law indices become $\alpha = 2$ and $\beta = -1/2$ (since 
 $M\propto T^{3/2}_{\rm vir}$). Our specific choice of $\alpha$ and $\beta$ for our fiducial model is purely illustrative, and allows 
 a strong mass-dependent scaling for the feedback limited galaxies, with a flatter slope contribution from the high-mass galaxies. 
 In the local Universe, observations show that for low-mass galaxies the star formation efficiency may increase with host halo mass, 
 $f_{\star} \propto M^{2/3}$ \citep{Kauffmann:2003p1701}. On the other hand, our framework also allows small-mass galaxies to 
 have even higher ionizing efficiencies than the high-mass galaxies. Such models \citep[e.g.][]{Alvarez:2012p3266} could be motivated 
 by a much higher escape fraction in small galaxies \citep[e.g.][]{Wise:2014p3267}. For our remaining \eor{} parameters, 
 we adopt $\zeta_{0}$ = 70, ${\rm log}_{10}T^{\rm Feed}_{\rm vir} = 4.7$ and $R_{\rm mfp}=15$~Mpc.
  
Again, we provide an illustrative example of our generalized \eor{} model by constructing a schematic picture of the expected
$z=8$ UV luminosity function in Fig.~\ref{fig:UVLum_DPL} following the same approach as in Section~\ref{sec:21CMFAST}.
We convert our new critical feedback temperature, $T^{\rm Feed}_{\rm vir}$ into a UV magnitude (total halo mass) of 
$M_{\rm UV} = -13.7$ ($M\sim10^{9.11}\,M_{\sun}$), and denote this and the atomic cooling threshold by the vertical dashed lines.
 For the massive (bright) galaxies, denoted by $\beta = -1/3$, we use the same fits to the UV luminosity function in 
 \citet{Kuhlen:2012p1506}. As before, this fit for these bright galaxies can be translated to $f_{\rm esc} \propto M^{-0.18}$ 
 (see the discussion in Section~\ref{sec:UVLum}).

\begin{table*}
\begin{tabular}{@{}lccccccccc}
\hline
Instrument & & & Parameter & & & & $\bar{x}_{\hi{}}$ &  \\
(multi-$z$) & $\zeta_{0}$ & $R_{\rm mfp}$~(Mpc) & $\alpha$ & $\beta$ & log$_{10}$$(T^{\rm Feed}_{\rm vir})$ & $z = 8$ & $z = 9$ & $z = 10$\\
\hline
\vspace{0.8mm}
HERA 331 & 68.28$\substack{+20.27 \\ -25.96}$ & 14.99$\substack{+2.89 \\ -2.36}$ & 1.10$\substack{+1.28 \\ -0.72}$ & -0.31$\substack{+0.39 \\ -0.45}$ & 4.74$\substack{+0.42 \\ -0.22}$ & 0.28$\substack{+0.04 \\ -0.03}$ & 0.55$\substack{+0.04 \\ -0.04}$ & 0.74$\substack{+0.04 \\ -0.04}$ \\
\vspace{0.8mm}
SKA & 70.20$\substack{+18.62 \\ -25.68}$ & 14.87$\substack{+2.79 \\ -2.13}$ & 1.11$\substack{+1.33 \\ -0.70}$ & -0.33$\substack{+0.37 \\ -0.43}$ & 4.73$\substack{+0.44 \\ -0.21}$ & 0.28$\substack{+0.03 \\ -0.03}$ & 0.55$\substack{+0.03 \\ -0.03}$ & 0.74$\substack{+0.03 \\ -0.03 }$ \\
\hline
\end{tabular}
\caption{Summary of the median recovered values (and associated 16th and 84th percentile errors) for the 
five reionization parameters, $\zeta_{0}$, $R_{\rm mfp}$, $\alpha$, $\beta$ and $T^{\rm Feed}_{\rm vir}$ from the 
generalized \eor{} model (two ionizing galaxy populations) and the associated \igm{} neutral fraction, $\bar{x}_{\hi{}}$. 
We assume a total 1000 h integration time for each of the three different epoch observations of the 21 cm PS at $z=8$, 9 and 10 
for HERA and SKA. Our fiducial mock observation corresponds to 
($\zeta_{0}$, $R_{\rm mfp}$, $\alpha$, $\beta$, ${\rm log}_{10}T^{\rm Feed}_{\rm vir}$) = (70, 15~Mpc, 4/3, --1/3, 4.7) which
results in an \igm{} neutral fraction of $\bar{x}_{\hi{}} = 0.28$, 0.55, 0.74 at $z=8$, 9 and 10 respectively.}
\label{tab:5param_summary}
\end{table*}

\begin{figure*} 
	\begin{center}
		\includegraphics[trim = 0.2cm 0cm 0cm 0.5cm, scale = 0.88]{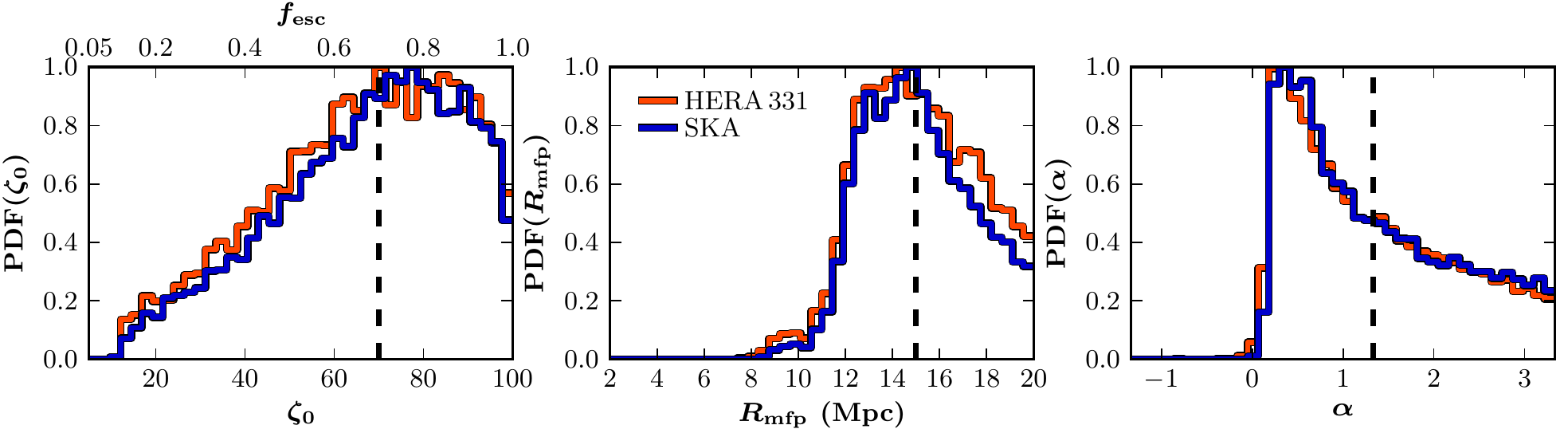}
		\includegraphics[trim = 0.2cm 0.5cm 0cm -0.2cm, scale = 0.88]{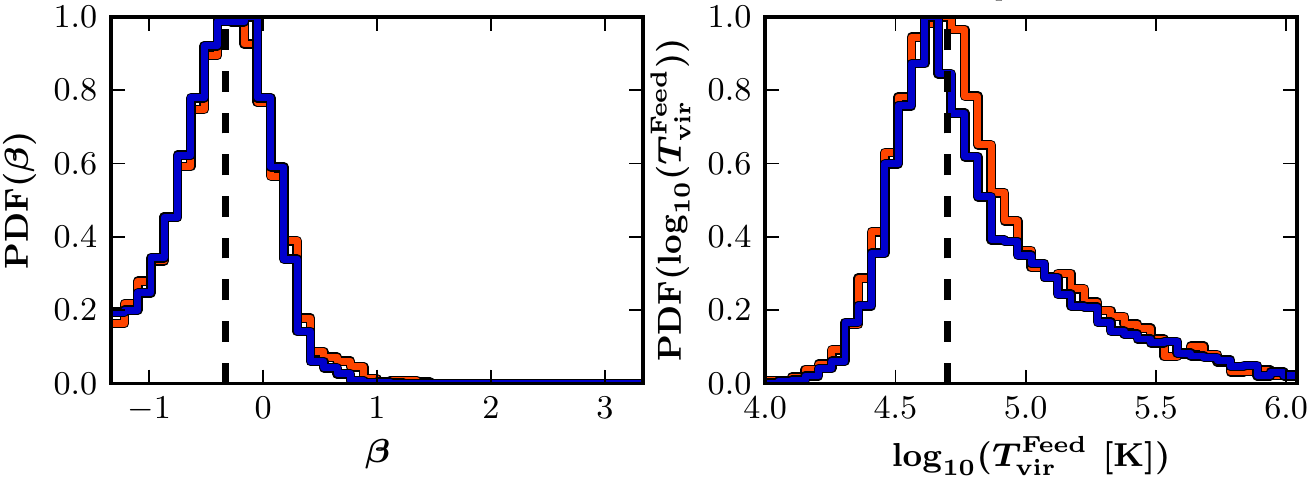}
	\end{center}
\caption[]
{The marginalized 1D PDFs for each of the five reionization parameters 
($\zeta_{0}$, $R_{\rm mfp}$, $\alpha$, $\beta$ and $T^{\rm Feed}_{\rm vir}$) from our generalized \eor{} model containing two 
distinct ionizing galaxy populations for a combined 1000 h observation of the 21 cm PS across multiple redshifts ($z=8,9$~and~10). We consider two separate telescope designs, HERA (red) and the SKA (blue), and the vertical lines denote our fiducial input parameters ($\zeta_{0}$, $R_{\rm mfp}$, $\alpha$, $\beta$, $T^{\rm Feed}_{\rm vir}$) = (70, 15~Mpc, 4/3, --1/3, 4.7).}
\label{fig:1DPDFs_HERA_SKA_DPL}
\end{figure*}

\subsection{Constraints from combining multiple redshift observations}

\begin{figure*} 
	\begin{center}
		\includegraphics[trim = 0.1cm 0.3cm 0cm 0.5cm, scale = 0.88]{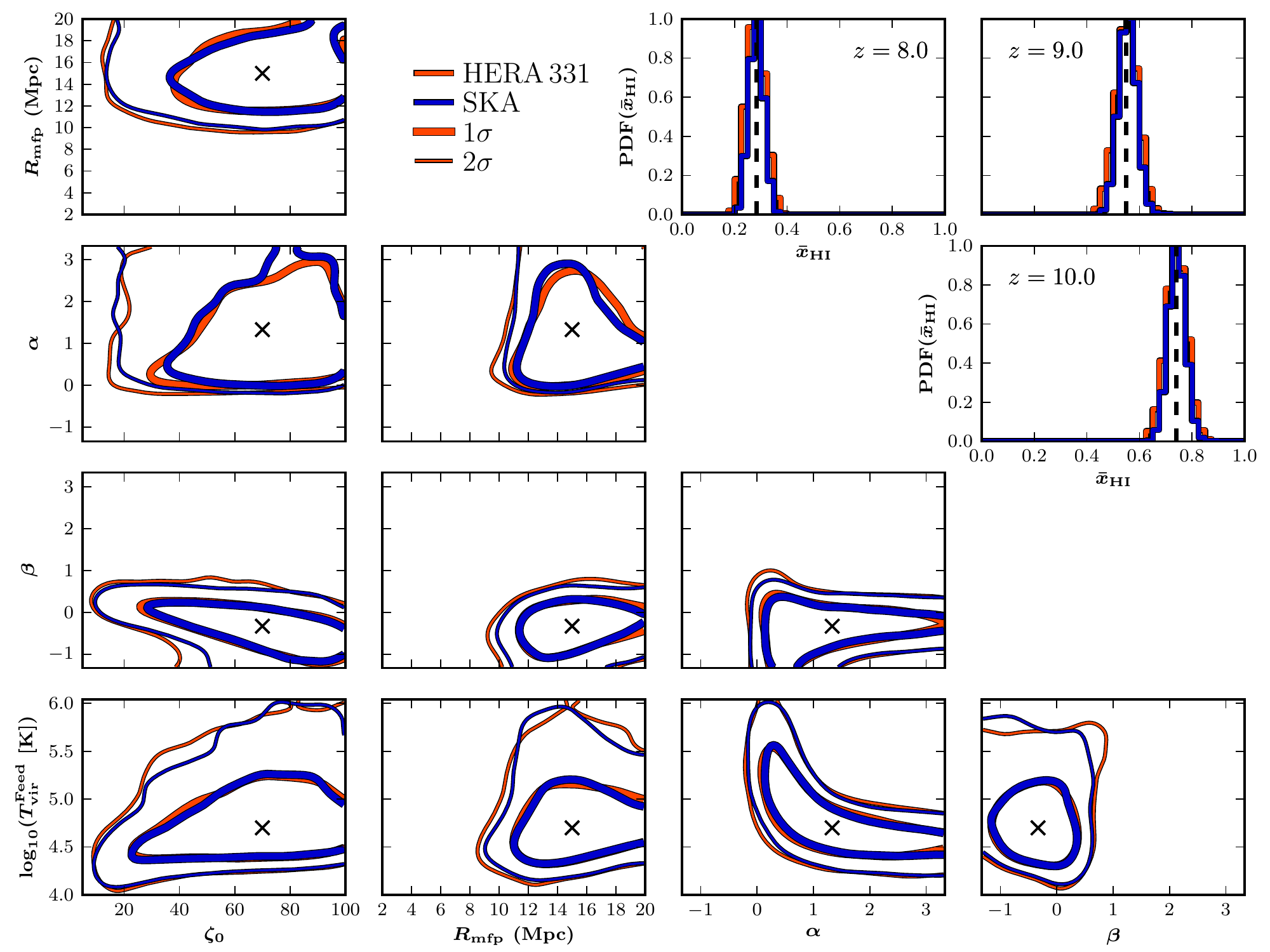}
	\end{center}
\caption[]
{The 2D joint marginalized likelihood contours for our generalized five parameter \eor{} model ($\zeta_{0}$, 
$R_{\rm mfp}$, $\alpha$, $\beta$ and $T^{\rm Feed}_{\rm vir}$) for a combined 1000 h observation of the 21 cm PS across 
multiple redshifts ($z=8,9$~and~10). We consider two separate telescope designs, HERA (red) and SKA (blue), crosses denote 
the fiducial model parameters and the 1$\sigma$ and 2$\sigma$ contours are shown as the thick and thin contours respectively. 
Additionally, in the top-right corner we provide the one-dimensional PDFs of the recovered \igm{} neutral fraction for each 
individual redshift observation, with the vertical dashed line denoting the neutral fraction of the mock 21 cm PS observations, 
$\bar{x}_{\hi{}} = 0.28$, 0.55 and 0.74 respectively. }
\label{fig:2D_HERA_SKA_DPL}
\end{figure*}

Given the increased parameter degeneracies in this generalized five parameter \eor{} model, we will focus only on 
multiple epoch observations of the 21 cm PS. As in Section~\ref{sec:multiz}, we assume three independent 1000 h observations 
of the 21 cm PS at $z=8$, 9 and 10; however, only for HERA and the SKA. In Fig.~\ref{fig:1DPDFs_HERA_SKA_DPL} we 
show the recovered 1D marginalized PDFs for each of the five \eor{} model parameters from HERA (red curves) and SKA 
(blue curves) and in Fig.~\ref{fig:2D_HERA_SKA_DPL} we show the corresponding 2D joint marginalized likelihood contours 
and the resulting 1D PDFs of the \igm{} neutral fraction at each observed epoch. Finally, in Table~\ref{tab:5param_summary}, 
we report the recovered median values and the corresponding 16th and 84th percentiles of our \eor{} model parameters. 

For the majority of these \eor{} parameters, we recover rather broad PDFs, noting mild to strong asymmetries emphasizing 
the strength of the degeneracies.  This is to be expected with more model parameters. As an example, focusing on the 
$\alpha$-${\rm log}_{10}T^{\rm Feed}_{\rm vir}$ panel for a decreasing $T^{\rm Feed}_{\rm vir}$ (increasingly narrower mass 
range for the feedback limited galaxies), any strong suppression (high $\alpha$) can mimic the same behaviour as our fiducial 
model. This is due to the fact that for our assumed value of $\alpha = 4/3$ the strong mass suppression in the ionizing 
efficiency only allows low-mass galaxies closest to $T^{\rm Feed}_{\rm vir}$ to provide any meaningful contribution to reionization. 

Despite these degeneracies, we are still able to recover meaningful constraints on these EoR parameters. For example, the 
contribution of low-mass galaxies to reionization as well as a well-defined feedback scale are picked up in the marginalized 
1D $T^{\rm Feed}_{\rm vir}$ PDFs.  Such fundamental issues in galaxy formation are unlikely to be addressed with direct 
observations, even with JWST.

\begin{figure*} 
	\begin{center}
		\includegraphics[trim = 0.2cm 0cm 0cm 0.5cm, scale = 0.88]{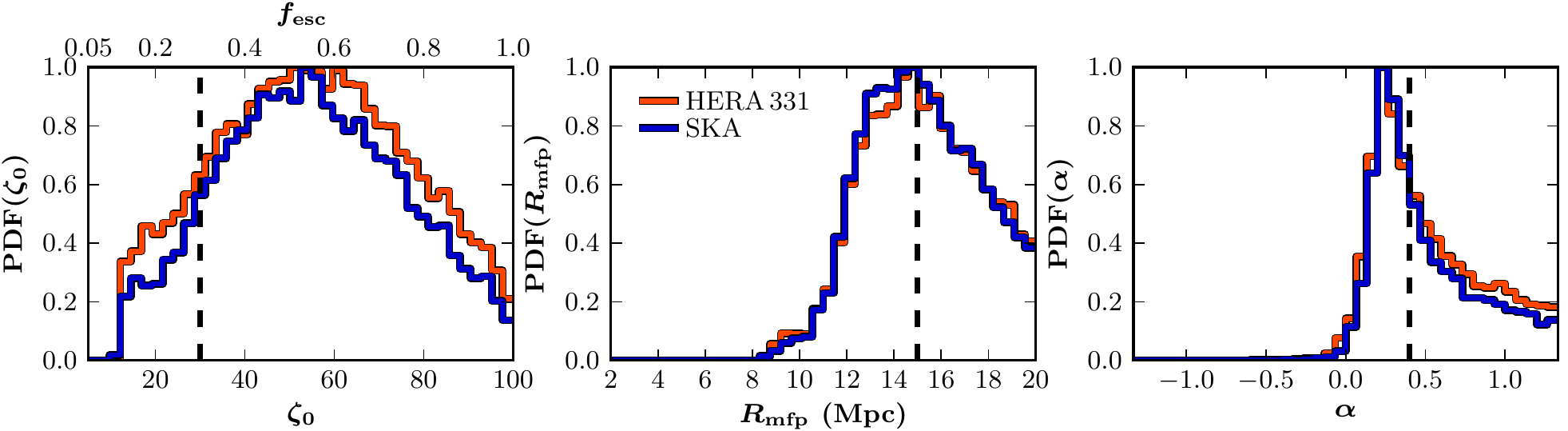}
		\includegraphics[trim = 0.2cm 0.5cm 0cm -0.2cm, scale = 0.88]{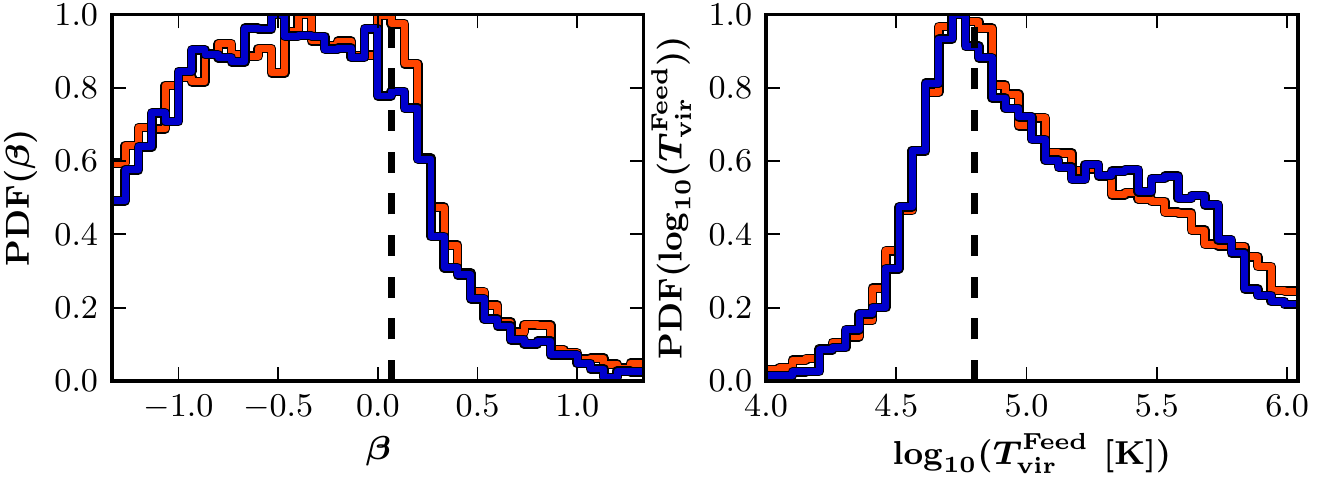}
	\end{center}
\caption[]
{Same as Fig.~\ref{fig:1DPDFs_HERA_SKA_DPL}, except now we consider an \eor{} model with a weaker break in the power-law 
indices. For this model, our \eor{} parameters are ($\zeta_{0}$, $R_{\rm mfp}$, $\alpha$, $\beta$, log $T^{\rm Feed}_{\rm vir}$) 
= (30, 15~Mpc, 2/5, 1/15, 4.8). As expected, reducing the amplitude of each power-law index and weakening the relative break
between them, reduces the overall constraining power on all \eor{} model parameters.}
\label{fig:1DPDFs_HERA_SKA_DPL_alternative}
\end{figure*}

Quantitatively, both HERA and the SKA are able to recover median values for the five \eor{} parameters to within 5 per cent
of their fiducial values with $\alpha$ being the exception ($\sim20$~per cent). The corresponding percentile errors for $\alpha$ 
are additionally rather broad, however, this was to be expected from the qualitative analysis above (i.e.\ a strong mass 
dependence of $\alpha$ is hard 
to constrain due to the narrow range in mass of these galaxies which can contribute to reionization). Finally, from 
Fig.~\ref{fig:2D_HERA_SKA_DPL}, we see that the PDFs of the \igm{} neutral fraction are tightly centred around the expected 
values from our fiducial model, and their quantitative errors are equivalent to our simple three parameter \eor{} model. This 
indicates that we have extracted all the available information from sampling the reionization history for both our 21 cm 
experiments. In principle, we could improve these constraints by (i) adding measurements of the 21 cm PS in additional bands 
(e.g.\ at $z=7$); (ii) applying priors on to our \eor{} model parameters; (iii) considering alternative
statistics of the 21 cm signal in combination with the 21 cm PS; or (iv) reducing the intrinsic modelling uncertainty.

It is important to note, that the discussion of the degeneracies and constraints of these \eor{} parameters is specific to our chosen 
fiducial model for the mock observation. Due to the sharp break in the power-law indices describing the two populations of ionizing 
galaxies and our choice for $T^{\rm Feed}_{\rm vir}$, the contribution to reionization from the high-mass (efficient star-forming) 
galaxy population can be relatively well constrained (as evident by the recovered PDF for $\beta$). Such a model is motivated 
by theoretical predictions of a sharp feedback scale \citep[e.g.][]{Finlator:2011p3408,Raicevic:2011p3440}. If however, we select 
a shallower break in the power-law indices, we effectively spread out the allowed range of masses of the ionizing galaxies 
(i.e.\ a shallower, positive $\alpha$ allows an increased contribution from the lower mass galaxies as they are not as 
efficiently suppressed). To illustrate this pessimistic case, in Fig.~\ref{fig:1DPDFs_HERA_SKA_DPL_alternative} we consider 
($\zeta_{0}$, $R_{\rm mfp}$, $\alpha$, $\beta$, $T^{\rm Feed}_{\rm vir}$) = (30, 15~Mpc, 2/5, 1/15, 4.8). As expected, we now 
recover much broader PDFs. Most notably, with the shallower break in the power-law indices, it becomes increasingly difficult to 
constrain $T^{\rm Feed}_{\rm vir}$ due to the smoother transition between the ionizing efficiency of the two galaxy populations. 
This is understandable, as in this model the two galaxy populations resemble a single population.

\section{Conclusion} \label{sec:conclusion}

The \eor{} is rich in astrophysical information, probing the nature of the first stars and galaxies, as well as their impact on the \igm{}.
However, our current understanding of EoR astrophysics is poorly understood. Our understanding will rapidly improve in the near 
future with the detection of the cosmic 21 cm signal from a broad range of dedicated 21 cm experiments. However, we are faced 
with the fundamental question, \textit{what exactly can we learn from these observations?}

To this end, we developed \cmmc{}\footnote{http://homepage.sns.it/mesinger/21CMMC.html}, a new, massively parallel MCMC 
analysis tool specifically designed to recover constraints on any astrophysical parameters during the \eor{} from observations of 
the cosmic 21 cm signal. \cmmc{} utilizes existing astrophysical simulation codes and parameter estimation techniques to ensure 
its efficiency and flexibility. It combines the astrophysical parameter sampler \CH{} \citep{Akeret:2012p842}, calling the 
seminumerical 21 cm simulation code \cmfst{} \citep{Mesinger:2007p122,Mesinger:2011p1123} at each point in the sampled \eor{} 
astrophysical parameter space. This allows sampling over $10^{5}$ individual realizations of the cosmic 21 cm signal to recover 
robust \eor{} parameter constraints, without common assumptions of Gaussian errors.

Throughout this work, we have emphasized several key strengths of \cmmc, outlining its applicability to a broad range of studies 
of the 21 cm signal during the \eor{}. In summary, we 
\begin{itemize}
\item[\textit{i})] showed by utilizing a Bayesian MCMC framework, we can efficiently and accurately recover robust astrophysical 
constraints by recovering the marginalized 1D PDFs, irrespective of any model degeneracies. This removes the fundamental 
limitations and assumptions inherent in previous \eor{} astrophysical parameter studies such as sampling fixed grids 
\citep[e.g.][]{Choudhury:2005p2859,Barkana:2009p116,Mesinger:2012p1131,Zahn:2012p1156}, 
Fisher Matrix approaches \citep{Pober:2014p35} or simplistic analytic modelling of the \eor{} 
\citep{Harker:2012p2856,Morandi:2012p2857,Patil:2014p2858};
\item[\textit{ii})] highlighted that it is generalizable to any theoretical model of the \eor{}. We showcased two such models: 
(i) a popular, three parameter \eor{} model driven by a single population of star-forming galaxies characterized by a constant 
(mass-independent) ionizing efficiency; and (ii) a generalized five parameter \eor{} model containing two ionizing galaxy 
populations each characterized by a unique mass-dependent power-law ionizing efficiency;
\item[\textit{iii})] highlighted the available constraining power from both single redshift and combining multiple redshift
observations of the ionization 21 cm PS for three current and proposed 21 cm experiments, LOFAR, HERA and the SKA;
\item[\textit{iv})] recovered for LOFAR/HERA/SKA fractional errors of 45.3/22.0/16.7, 33.5/18.4/17.8 and 6.3/3.3/2.4 per cent 
on the ionizing efficiency (escape fraction), mean free path of ionizing photons and the minimum halo mass hosting star-formation 
from combining three independent 1000hr observations of the 21 cm PS at $z=8$, 9 and 10 assuming no priors and a 
conservative noise estimate;
\item[\textit{v})] showed that by removing the EoR modelling uncertainty from our fiducial analysis, these constraints can be improved by up to a factor of a $\sim$few.  This result motivates further work on characterizing EoR modelling errors in order to maximize science returns from second-generation instruments;
\item[\textit{vi})] quantified the improvements in constraining power on the \eor{} parameters from accessing
the additional information from the cosmic reionization history through the application of physically motivated priors on the 
\igm{} neutral fraction at the tail-end of reionization;
\item[\textit{vii})] emphasized that \cmmc{} additionally allows for the inclusion of individual priors on the \eor{} model parameters 
motivated by current and upcoming observations and theoretical advances. Furthermore, while we have focused solely on the 
21 cm PS, it is easily adaptable to consider any statistical measure of the cosmic 21 cm signal.
\end{itemize}

Finally, due its applicability to a broad range of \eor{} studies, \cmmc{} could be an important analysis tool for the 21 cm \eor{} 
community. For example, it can be used to quantify how foregrounds and other contaminants can hinder the recovery of
EoR astrophysics. Moreover, it can serve to guide designs and observing strategies for current and future 21 cm experiments to 
maximize their scientific returns.

\section*{Acknowledgements}

We thank the anonymous referee for their helpful suggestions.
We are grateful to Jonathan Pober, Adrian Liu and Adam Lidz for valuable comments on a draft version of this manuscript.  Additionally, we thank Jonathan Pober for making his telescope sensitivity code \sense{} publicly available and also
Joel Akeret and Sebastian Seehars for the development and public release of \CH{}.

\bibliography{21cmMC}

\end{document}